\documentclass{jfm}

\usepackage{graphicx}
\usepackage{epstopdf, epsfig}
\usepackage{enumerate}
\usepackage{amsmath}
\usepackage{amsfonts}
\usepackage{amssymb}
\usepackage{mathrsfs}
\usepackage[breaklinks, colorlinks, citecolor=blue]{hyperref}
\usepackage{color}
\usepackage{float}
\usepackage[english]{babel}
\usepackage{txfonts}
\usepackage{url}
\usepackage{array}
\usepackage{RefJ}
\usepackage{empheq}

\newcommand*\dd{\mathop{}\!\mathrm{d}}
\renewcommand{\vec}[1]{\mathbf{#1}}

\shorttitle{Gravitational instability in self-gravitating planar polytropes}
\shortauthor{J.-B. Durrive and M. Langer}

\title{Analytic growth rate of gravitational instability in self-gravitating planar polytropes}

\author{Jean-Baptiste Durrive\aff{1}
  \corresp{\email{jean.baptiste.durrive@e.mbox.nagoya-u.ac.jp}}
 \and Mathieu Langer\aff{2}}

\affiliation{\aff{1}Department of Physics and Astrophysics, Nagoya University, Nagoya 464-8602, Japan
\aff{2}Institut d'Astrophysique Spatiale, CNRS, UMR 8617, Univ. Paris-Sud, Universit\'e Paris-Saclay, B\^at. 121, 91405 Orsay, France}

\begin{document}

\maketitle

\begin{abstract}
Gravitational instability is a key process that may lead to fragmentation of gaseous structures (sheets, filaments, haloes) in astrophysics and cosmology. We introduce here a method to derive analytic expressions for the growth rate of gravitational instability in a plane stratified medium. We consider a pressure-confined, static, self-gravitating fluid of arbitrary polytropic exponent, with both free and rigid boundary conditions. The method we detail here can naturally be generalised to analyse the stability of more complex systems.
Our analytical results are in excellent agreement with numerical resolutions.
\end{abstract}


\section{Introduction}

Sheets and filaments of matter appear in many different astrophysical contexts. In the interstellar medium of galaxies, the sheet-like and filamentary structure of giant molecular clouds has been known for a long time. It results from the conspiring action of supernova explosions, thermal instability, cloud-cloud collisions, turbulence, and magnetic fields \citep[e.g.][]{Schneider1979,Bally1987,Mizuno1995, Hartmann2002,Myers2009,Pudritz2013,Andre2014,Andre2015,Federrath2016,Kalberla2016}. In the intergalactic medium, numerical simulations demonstrate that matter gets gravitationally organised into a cosmic web of voids delineated by cosmological walls and filaments \citep[e.g.][]{Klypin1983,Klar2010}. The nodes of the web, hosting galaxies and galaxy clusters, are supplied with matter, baryonic and dark, flowing along the filaments that interconnect them. Part of this accretion occurs intermittently \citep[e.g.][]{Keres2009a, Dekel2009,Dekel2009b,SanchezAlmeida2014}, suggesting that denser clumps of matter might form not only within the nodes of the cosmic web, but also in either voids, walls or filaments. It has been pointed out that a fraction of the clumps may be of artificial origin due to numerical effects that are inherent to classical Smooth Particle Hydrodynamics numerical codes \citep[see][for a discussion]{Hobbs2013, Nelson2013}, and less present in simulations based on moving mesh techniques \citep[cf.][]{Springel2010}. However, the rest of the clumps most probably has a true physical origin \citep{Hobbs2016}. Are the clumps in filaments and cosmic walls observed in cosmological numerical simulations solely the product of the growth of primordial overdensities? Are these gas clumps always subtended by collapsed dark matter haloes, or is it possible that baryon fragments form and grow thanks to (sub-grid) gravitational instability? The fragmentation of self-gravitating sheet-like and filamentary structures may in principle occur through many different instabilities. In the cosmological context, thermal, Rayleigh-Taylor, Kelvin-Helmholtz, etc., may play a r\^ole in the denser environments of massive haloes \citep[e.g.][]{Keres2009b}. In the more dilute environment of the filamentary cosmic web, gravity is the universal actor at play.

In order to answer these questions, we need to understand fully the gravitational fragmentation of the baryonic fluid. A full analytic treatment of gravitational instability in inhomogeneous, continuous media is still missing in the literature. Formally, the study of gravitational instability boils down to an eigenvalue problem that is not easy to solve because the corresponding system of equations is of fourth order, with complicated coefficients. Historically, the investigation of  gravitational instability was triggered by the work of Jeans \citep[e.g.][]{Jeans28}. Other seminal works include \citet{Ledoux1951} and \citet{Simon1965} for sheet-like structures, and \citet{Chandrasekhar1953} who explored the cylindrical case of isothermal magnetised filaments, essentially in the context of the interstellar medium. These were then followed by many studies, exploring further the r\^ole of the various ingredients relevant to describe astrophysical and cosmological environments, notably the presence of an external pressure \citep[e.g.][]{Elmegreen1978, MiyamaEtAl87a,MiyamaEtAl87b,Narita1988},  uniform or differential rotation \citep[e.g.][]{Safronov1960, Simon1965b, Narita1988, Burkert2004},  flow \citep[e.g.][]{Lacey89},  the background expansion of the Universe and the dark matter component in the cosmological context \citep[e.g.][]{Umemura93,AnninosEtAl95,HosokawaEtAl00}, the possible advent of convective instability \citep[e.g.][]{MamatsashviliRice10,BreysseEtal14}, the local expansion (or collapse) of the structure \citep[e.g.][]{InutsukaMiyama92,IwasakiEtAl2011}, etc. And of course, a lot of focus has been put on magnetic fields, given their importance in interstellar environments \citep[e.g.][]{Strittmatter1966,Kellman1972,Kellman1973,Langer1978,Nakano1978,TomisakaIkeuchi83,Nakano1988,HosseiniradEtAl17}. These studies were performed in either planar or cylindrical geometries.

In this paper, we concentrate our attention exclusively on the r\^ole of gravity. We study analytically the stability of a planar, pressure-confined, static, self-gravitating, polytropic fluid. We focus on the planar geometry, leaving the cylindrical case for future work, because the fragmentation of planar structures is the first key step in the full process of fragmentation as suggested for instance by studies of linear and non-linear growth of perturbations (e.g.\ in isothermal sheets in \citet{MiyamaEtAl87a,MiyamaEtAl87b}) which show that gas layers fragment into numerous filaments which subsequently fragment into small blobs that ultimately constitute the star-forming regions \citep[see][]{Larson85, InutsukaMiyama97}. In addition, from a formal point of view, cylindrical geometry adds a couple more difficulties that are all the better figured out once the planar case is clear.
Then, among the numerous physical ingredients that are relevant to the astrophysical and cosmological contexts,  we consider here a pressure-confined structure with the two different types of boundary conditions commonly used in the literature for their relevance and generality. Most critically, we keep the polytropic exponent $\gamma$ arbitrary, while a majority of authors so far focused on the special isothermal case $\gamma=1$ because it has the property of corresponding to a stratified equilibrium density with a simple analytic expression and with a uniform speed of sound, which simplifies greatly the equations governing the dynamics. In that sense, our results are already quite rich in terms of physics. Most importantly, it is very natural and relatively straightforward to generalise the method we introduce here to include the aforementioned additional physical ingredients. Increasing the complexity progressively will help disentangle the r\^ole of each element in the dynamics, and we leave this effort for future work.

In the papers mentioned above, a majority of authors solve the equations and obtain dispersion relations numerically. Others like \citet{VanLooEtAl14} or \citet{DinnbierEtAl17} make use of numerical simulations. Analytic results were derived in special cases only, e.g. in the incompressible case \citep[e.g.][]{GoldreichLyndenBell1965,Tassoul67} or the thin sheet limit \citep[e.g.][]{TomisakaIkeuchi85,WunschEtAl10}, or with restrictive scope like focusing on marginal stability \citep[computing the critical wavenumber but not the maximum growth rate and  wavenumber, e.g.][]{Oganesyan60,GoldreichLyndenBell1965}, or working under simplifying assumptions about the scale of perturbations \citep[e.g.][]{LubowPringle93,Clarke99}. A number of authors examine this problem through variational approaches \citep[e.g.][]{Chandrasekhar1961,LyndenBellOstriker67,RaoultPellat78}. They provide general stability criteria but do not give explicit expressions for the eigenvalues. Recently, \citet{KeppensDemaerel16PartII} derived an upper bound on the perturbed self-gravitational energy associated with the Lagrangian displacement.
Here, we are interested in deriving analytic expressions of the growth rate as a function of the transverse wavenumber. As far as planar pressure-confined self-gravitating polytropes are concerned, to the best of our knowledge, the most analytical studies are the seminal work of \citet{GoldreichLyndenBell1965}, and a recent contribution from \cite{KimEtAl12} which both include considerations about rotation. In \citet{GoldreichLyndenBell1965} the authors consider a generic polytropic exponent $\gamma$, but ultimately derive explicit dispersion relations only in the incompressible case, and then for $\gamma=1$ and $\gamma=2$ they derive the expressions of the critical wavenumber by solving the equations for the growth rate $\omega$ near zero. \cite{KimEtAl12} however explicit analytic expressions approximating the numerical dispersion relation for an arbitrary value of $\gamma$. They provide very good approximations with very simple expressions, but their derivation is unfortunately not systematic, in the sense that their final result relies on an assumption about the shape of the perturbation and on numerical fits that allow them to essentially guess the approximate functional dependence of the fundamental frequency on the transverse wavenumber. It seems therefore difficult to improve the accuracy of their results and to generalise them in order to include more physics.

Here, we introduce a method that allows us to obtain explicitly the dispersion relation between the fundamental frequency and the transverse wavenumber in terms of the physical properties of the structure in principle up to arbitrary precision in a systematic way. 
For that purpose, we reformulate and decompose the full fourth-order problem into a sequence of second-order problems that can be solved separately. Interestingly, we show that the potentially unstable fundamental frequency can be obtained from the stable higher order harmonics. Therefore, while the aim is to compute the unstable part of the spectrum,  we will also provide new expressions for its stable part as a by-product. The paper is organised as follows. In section \ref{section:GoverningEquations} we present the equations governing the equilibrium state and the perturbations which, together with the boundary conditions, bring us to formulate the eigenvalue problem. In section \ref{section:Reformulation} we reformulate these equations into a form that is more convenient for expressing the eigenvalue equation, i.e. a scalar equation whose solutions are the eigenvalues. Then, in section \ref{section:Method}, we compute the fundamental eigenfrequency, corresponding to the growth rate of the gravitational instability. For clarity, we decomposed this section into three steps, and we show in the main text only the key intermediate results leading to formula \eqref{omega02_coeffs_explicit}, which is the main explicit result of this paper. We provide the somewhat intricate details of the underlying derivations in appendices \ref{section:StepI} and \ref{section:StepII} for steps I and II respectively. We conclude in section \ref{section:Conclusion}.

\section{Governing equations}
\label{section:GoverningEquations}

In this section, we first present the equations governing the equilibrium state, then the linearised equations of motion satisfied by the perturbations, and finally the boundary and symmetry conditions considered. This ultimately brings us, in section \ref{section:EigenvalueProblem}, to formulate the eigenvalue problem at the heart of this stability analysis. For now we present these equations in the form usually found in the literature, but in section \ref{section:Reformulation} we reformulate them in a form that is more efficient for our purpose.

\subsection{Equilibrium state}
\label{section:EquilibriumState}

\begin{figure}
\centering
\includegraphics[scale=0.6]{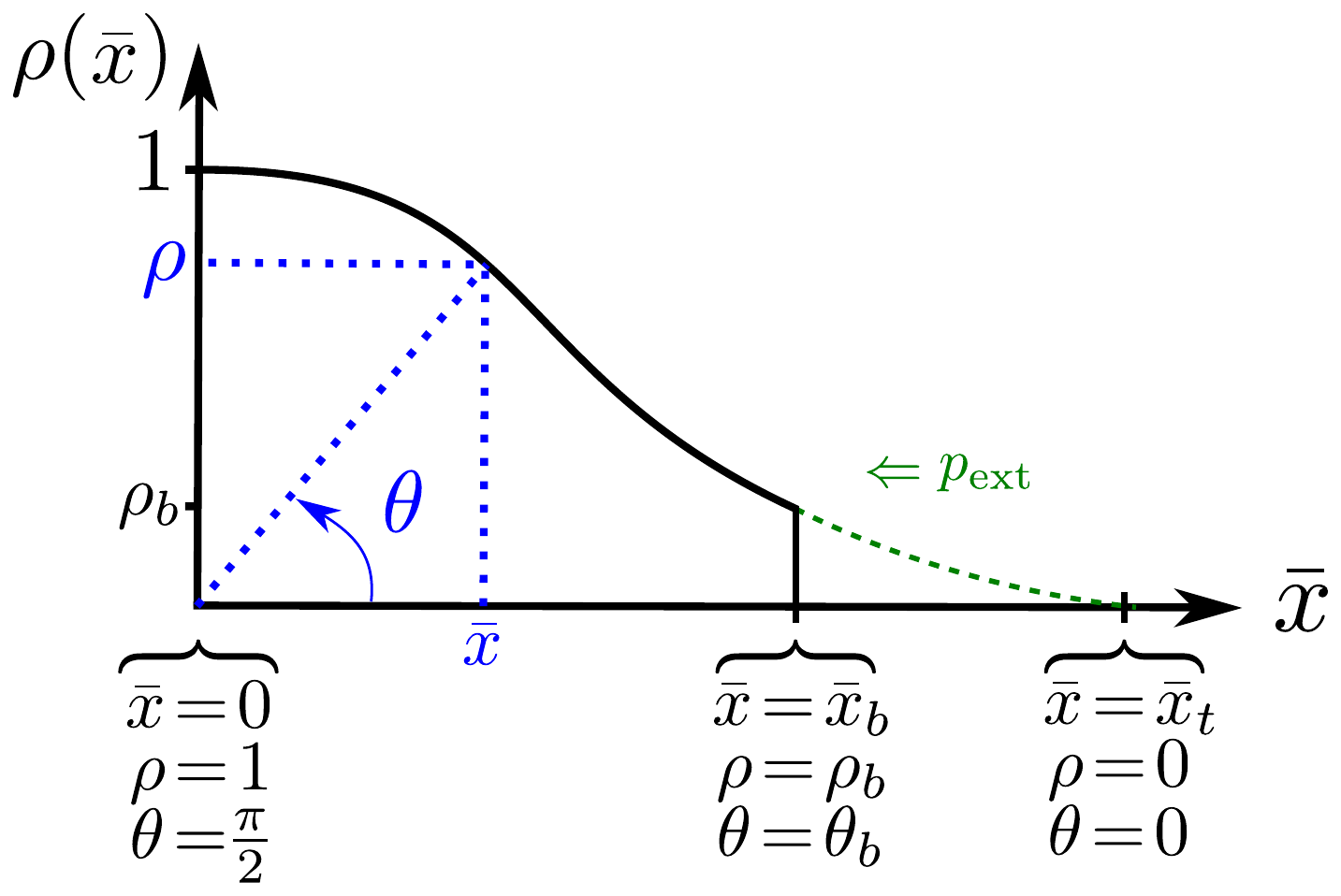}
\caption{Sketch illustrating the important quantities of (i)~The equilibrium state: In black, the typical density profile (normalised density $\rho$ as a function of normalised position $\bar{x}$) of a self-gravitating, planar polytrope. It is confined by an external pressure $p_\mathrm{ext}$ represented by the green arrow. The boundary is at the position $\bar{x}_b$ at which the internal pressure equates $p_\mathrm{ext}$. The dashed green line indicates how the profile would look like if $p_\mathrm{ext}$ were equal to zero. The position $\bar{x}_t$ is such that beyond it the solution to the Lane-Emden equation becomes unphysically negative.
(ii)~The perturbed state: In this paper we describe perturbations with the variables $\rho$ and $\theta$ rather than $\bar{x}$. These are indicated in blue. The blue arrow indicates qualitatively how $\theta$ evolves as $\bar{x}$ decreases and $\rho$ increases. Keep in mind however that this figure is really just an illustration: the behaviour indicated by the blue arrow is only qualitative, and it should not be read geometrically as $\tan \theta = \rho / \bar{x}$. See definitions \eqref{ChangeVariable_xRho} and \eqref{ChangeVariable_RhoTheta} for their precise relation.}
\label{fig:SchemaPerturbedProfile}
\end{figure}

In figure~\ref{fig:SchemaPerturbedProfile} we sketch how the equilibrium density profile of a planar, pressure-confined, static, self-gravitating, polytropic structure typically looks like. We use the $x$-direction as the direction of stratification, while it is homogeneous with infinite extent in the $y$ and $z$ directions. Therefore all equilibrium quantities, marked with a subscript $0$, depend only on position $x$. The equilibrium state is assumed to be static, because it is methodologically convenient to start with this case before generalising to more realistic situations in future work, and because it is physically relevant for systems with slow accretion, typically slower than the growth of  perturbations. The hydrostatic equilibrium reads
\begin{equation}
- \vec{\nabla} p_0 + \rho_0 \vec{g}_0 = \vec{0}
\label{HydroEquil_VectorForm}
\end{equation}
where $\rho_0$, $p_0$ and $\vec{g}_0$ are respectively the equilibrium density, pressure and gravitational acceleration. The system is self-gravitating, meaning that its potential well is shaped by its own density only, which translates into the Poisson equation
\begin{equation}
\vec{\nabla} \cdot \vec{g}_0 = - 4 \pi G \rho_0
\label{PoissonEqn}
\end{equation}
where $G$ is Newton's constant. For our choice of closure relation, we proceed as follows. In the most general case, pressure and density are related by an equation of state of the form $p = p(\rho,s)$ or $p = p(\rho,T)$ where $s$ and $T$ are respectively the specific entropy and the temperature. However, for a non-magnetised fluid to be at rest in a gravitational field, it must necessarily be barotropic, i.e. $p = p(\rho)$. Indeed, taking the curl of the hydrostatic equilibrium \eqref{HydroEquil_VectorForm}, since the curl of a gradient vanishes and $\vec{g}_0$ is the gradient of the gravitational potential, we end up with $\vec{\nabla} \rho_0 \times \vec{\nabla} p_0 = \vec{0}$. This means that the  density and  pressure gradients are aligned everywhere, and surfaces of constant density  coincide with surfaces of constant pressure. Here we consider the case of a polytropic equation of state, 
\begin{equation}
p_0 = \kappa \rho_0^\gamma
\label{p0_polytropic}
\end{equation}
where $\gamma$ is the polytropic exponent and $\kappa$ is a constant. In this work we derive our results for an arbitrary polytropic index, which makes them very general.

For a polytrope \eqref{p0_polytropic}, the adiabatic speed of sound $c_a^2 \equiv \frac{\partial p_0}{\partial \rho_0}$ is given by
\begin{equation}
c_a^2 = \kappa \gamma \rho_0^{\gamma-1}.
\label{ca2}
\end{equation}
Also, the hydrostatic equilibrium \eqref{HydroEquil_VectorForm} may be rewritten as
\begin{equation}
g_0 = c_a^2 \frac{\rho_0'}{\rho_0}.
\label{HydroEquilibrium}
\end{equation}
Plugging this in \eqref{PoissonEqn} gives the equation satisfied by the equilibrium density $\rho_0$ namely
\begin{equation}
(\rho_0^{\gamma-1})''+ \frac{4 \pi G}{\kappa} \frac{\gamma-1}{\gamma} \rho_0=0
\label{LaneEmden}
\end{equation}
which is known as the Lane-Emden equation.
Strictly speaking, we consider here $\gamma \neq 1$, but in fact all the results presented in this paper are valid for the isothermal case as well, by taking the limit $\gamma \rightarrow 1$. Doing the analysis taking $\gamma = 1$ from the beginning gives interesting analytical results, that we will explicit as part of a separate paper. As detailed in \cite{Horedt04}, the Lane-Emden equation for planar polytropes has simple analytic solutions only for the special cases $\gamma = \frac{2}{3}, 1, 2$ and $\infty$. However, more can be said  for instance on the extent of these structures: $\gamma \leq 1$ polytropes extend a priori to infinity, while for $\gamma > 1$ the solution has negative $\rho$ values, which are unphysical. For the latter, it is therefore common to define the width of the structure by the position $x_t$ at which the solution first vanishes. Multiplying \eqref{LaneEmden} by $(\rho_0^{\gamma-1})'$, integrating twice and using $\rho_0(x_t)=0$, one can show that it is given by
\begin{equation}
x_t = \sqrt{\frac{\kappa}{8 \pi G}} \rho_c^{\frac{\gamma}{2}-1} B\left(\frac{\gamma-1}{\gamma},\frac{1}{2}\right)
\label{Thickness}
\end{equation}
where $B$ is the Beta function and $\rho_c \equiv \rho_0(x=0)$ is the mid-plane value of the density. The above length $x_t$ must be taken with an important caveat too: by definition at this position the density vanishes, therefore the speed of sound also reaches zero, so that at this boundary any motion is supersonic and we would expect shock waves \citep{GoldreichLyndenBell1965}. In addition, the infinite extent of $\gamma \leq 1$ polytropes is not realistic. We may go around these two complications by improving the modeling a little, taking into account the existence of an external pressure $p_\mathrm{ext}$, which is relevant both in the astrophysical and cosmological contexts. For any $\gamma$, we truncate the solution of the Lane-Emden equation \eqref{LaneEmden} at the position $x_b$ such that the internal pressure equates the external one, i.e.
\begin{equation}
p_0(x_b) = p_\mathrm{ext}
\label{Def:xb}
\end{equation}
and this defines the boundary of the structure. We illustrate all this in figure \ref{fig:SchemaPerturbedProfile}. 

\subsection{Equations of motion}

Perturbations, marked with a subscript $1$, are described in terms of the Lagrangian displacement vector $\vec{\xi}$. In the case of a static equilibrium, the Eulerian velocity perturbation $\vec{v}_1$ is related to $\vec{\xi}$ by
\begin{equation}
\vec{v}_1 = \partial_t \vec{\xi}
\label{v1}
\end{equation}
so that here the linearised momentum conservation reads
\begin{equation}
\rho_0 \p_t^2 \vec{\xi} = \vec{F}(\vec{\xi})
\label{LinearizedMomentumConservation}
\end{equation}
with the so-called force operator \citep{Bernstein58,KeppensDemaerel16PartI}
\begin{equation}
\vec{F}(\vec{\xi}) = - \vec{\nabla} p_1 + \rho_1 \vec{g}_0 + \rho_0 \vec{g}_1
\label{ForceOperator}
\end{equation}
where $\rho_1$, $p_1$ and $\vec{g}_1$ are, respectively, the density, pressure and gravitational acceleration perturbations. In (\ref{ForceOperator}), the first term gives rise to sound waves due to the compressibility of the fluid. The second corresponds to the fact that the local density perturbation is evolving in an background potential well dictated by the equilibrium potential. Finally, the third term corresponds to the fact that perturbations generate  local potential wells which affect the entire density profile $\rho_0$ of the structure. This term is the source of Jeans' gravitational instability, which is the focus of our study here.

The linearised mass continuity equation expressed in terms of $\vec{\xi}$ is
\begin{equation}
\rho_1 = - \vec{\nabla} \cdot (\rho_0 \vec{\xi})
\label{LinearizedMassConservation}
\end{equation}
and the perturbation of the gravitational acceleration $\vec{g}_1$ satisfies the linearised Poisson equation
\begin{equation}
\vec{\nabla} \cdot \vec{g}_1 = - 4 \pi G \rho_1,
\label{LinearizedPoisson}
\end{equation}
keeping in mind that it is a curl-free vector field, namely
\begin{equation}
\vec{\nabla} \times \vec{g}_1 = \vec{0}.
\label{g1_CurlFree}
\end{equation}
It is only with that additional constraint that \eqref{LinearizedPoisson} is equivalent to the more usual form $\Delta \phi_1 = 4 \pi G \rho_1$, where $\phi_1$ is the perturbed gravitational potential, since $\vec{g}_1 = - \vec{\nabla} \phi_1$. 

In order to close this set of equations for the perturbations, we need an additional relation. Let us consider the case in which the timescale of the perturbations, i.e.\ the oscillation period if stable and growth time if unstable, is shorter than the timescale of the heat transfer between neighbouring fluid elements. The evolution of  perturbations may then be considered as adiabatic, and it can be shown \citep[e.g.][]{Thompson} that the equation expressing the absence of heat
$\delta Q=0$ translates into the following relation between the Lagrangian variation of pressure $\delta p$ and of density $\delta \rho$:
\begin{equation}
\frac{\delta p}{p_0} = \gamma_\mathrm{ad} \frac{\delta \rho}{\rho_0},
\label{AdiabaticPerturbLagrangian}
\end{equation}
where, in general, $\gamma_\mathrm{ad}$ is not equal to  the polytropic exponent $\gamma$ of the polytropic equation of state of the equilibrium \citep[e.g.][]{Cox80,Toci2015}. Here however, we will take them equal  because the purpose of this work is to focus on the gravitational instability of the Jeans type. This assumption switches-off buoyancy so that we get rid of convective instability, and as far as the stable part of the spectrum is concerned, since the Brunt-V\"ais\"al\"a frequency is now vanishing, the acoustic oscillations we will obtain are equivalent to the p-modes of stellar physics, without complicating the description with g-modes. Finally, rewriting \eqref{AdiabaticPerturbLagrangian} using the Eulerian variables $\rho_1$ and $p_1$ (related to the Lagrangian description by $\delta \rho = \rho_1 + \vec{\xi} \cdot \vec{\nabla} \rho_0$ and $\delta p = p_1 + \vec{\xi} \cdot \vec{\nabla} p_0$) yields our closure relation
\begin{equation}
p_1 = c_a^2\ \rho_1.
\label{p1_NoConvection}
\end{equation}

We look for solutions by separation of variables, i.e. we consider that all quantities $Q$ (namely $\xi_x, \xi_y, \xi_z, g_{1x}, g_{1y}, g_{1z}, \rho_1$ and $p_1$) are of the form $Q(x,y,z,t) = \hat{Q}(x)~Y(y)~Z(z)~T(t)$ since we are considering a planar geometry. Naturally, the respectively space- and time-translational invariance along the $y$, $z$ and $t$ dimensions yields exponential dependencies for $Y$, $Z$ and $T$. Also, given that we are considering a planar geometry with an infinite extent transverse to the stratified $x$-direction, the $y$ and $z$ directions are equivalent, so that we can rotate the axes in the ($y$, $z$) plane in order to remove any dynamics in the $z$-direction. Therefore, without loss of generality we take $k_z=0$ and we have
\begin{equation}
Q(x,y,z,t) = \hat{Q}(x) e^{i(k_y y - \omega t)}.
\label{Q_Fourier}
\end{equation}

\subsection{Boundary conditions and symmetry conditions}
\label{section:BC_firstFormulation}

The main difficulty to study gravitational instability is that the differential equations satisfied by the perturbations are of order four. This is due to the term $\vec{g}_1$ which, with \eqref{LinearizedPoisson}, turns the force operator \eqref{ForceOperator} into an integro-differential operator. Dealing with fourth-order equations, we need to impose four conditions. In this paper we consider boundary conditions~(BC) and symmetry conditions commonly used in the literature. We briefly summarise them here and adapt them to our notations. If need be, one may consult for example  \cite{GoldreichLyndenBell1965}, \cite{Elmegreen1978} or \cite{KimEtAl12} for more details.

Due to the reflection symmetry of the governing equations with respect to the $x=0$ plane,
the general solution is a superposition of symmetrical and antisymmetrical modes, which may be considered separately. It has been shown that antisymmetric modes are stable as far as gravitational instability is concerned. Therefore we will focus on symmetric modes, for which
\begin{equation}
\left\{
\begin{array}{l}
\hat{\xi}_x(x=0) = 0\\
\hat{g}_{1x}(x=0) = 0,
\end{array}
\right.
\label{SymmetryConditions}
\end{equation}
where we recall that quantities with a hat are defined through \eqref{Q_Fourier}. We will refer to (\ref{SymmetryConditions}) as the `symmetry conditions'.

Another condition is obtained as follows. Apply the divergence theorem to the linearised Poisson equation for an infinitesimally thin shell containing the boundary layer. Compute the gravitational acceleration outside the slab by solving Laplace's equation (since the external fluid is assumed to remain unperturbed), using the fact that it should not diverge at infinity. This results in the constraint
\begin{equation}
\hat{g}_{1x}(x_b) - i \hat{g}_{1y}(x_b) = 4 \pi G \rho_b \hat{\xi}_x(x_b).
\label{BC1_OldVariables}
\end{equation}

Finally, we consider two complementary types of BCs. First, the rigid BC corresponds to the case in which the surface at the boundary does not move, i.e. the Lagrangian displacement vanishes
\begin{equation}
\hat{\xi}_x(x_b) = 0  \hspace{1cm} \mathrm{(Rigid \ BC)}.
\label{BC2_Rigid_OldVariables}
\end{equation}
Second, on the contrary, we leave the boundary surface move freely, but require continuity of pressure throughout the dynamics, i.e. that the Lagrangian variation of pressure $\delta p = p_1 + \vec{\xi} \cdot \vec{\nabla} p_0$ vanishes at the boundary. With the hydrostatic equilibrium \eqref{HydroEquil_VectorForm}, the fourth condition in this Free BC case is thus
\begin{equation}
\hat{p}_1(x_b) = - \rho_0(x_B) g_0(x_b) \hat{\xi}_x(x_b) \hspace{1cm} \mathrm{(Free \ BC)}.
\label{BC2_Free_OldVariables}
\end{equation}

\subsection{The eigenvalue problem}
\label{section:EigenvalueProblem}

Our setting is constituted by the closed set of equations \eqref{LinearizedMomentumConservation}, \eqref{LinearizedMassConservation}, \eqref{LinearizedPoisson}, \eqref{g1_CurlFree}, and \eqref{p1_NoConvection}, together with the conditions \eqref{SymmetryConditions}, \eqref{BC1_OldVariables}, \eqref{BC2_Rigid_OldVariables} and \eqref{BC2_Free_OldVariables}. With the form \eqref{Q_Fourier} of the perturbations, this set becomes
\begin{equation}
\begin{array}{l}
\left\{
\begin{array}{l}
\hat{\rho}_1 = - (\rho_0 \hat{\xi}_x)' - \rho_0 i k_y \hat{\xi}_y\\
- \rho_0 \omega^2 \hat{\xi}_x = - c_a^2 \hat{\rho}_1' + \left(\hat{g}_0 - (c_a^2)'\right) \hat{\rho}_1 + \rho_0 \hat{g}_{1x}\\
- \rho_0 \omega^2 \hat{\xi}_y = - i k_y c_a^2 \hat{\rho}_1 + \rho_0 \hat{g}_{1y}\\
\hat{g}_{1x}' + i k_y \hat{g}_{1y} = - 4 \pi G \hat{\rho}_1\\
\hat{g}_{1x} = \left(\frac{\hat{g}_{1y}}{i k_y}\right)'
\end{array}
\right.
\end{array}
\label{EigenvalueProblem}
\end{equation}
where here a prime denotes differentiation with respect to $x$, $'=\frac{\dd}{\dd x}$. From top to bottom these are respectively: mass continuity, the $x$ and $y$ components of momentum conservation, Poisson equation and the curl-free condition. Together with the boundary and symmetry conditions, this constitutes an \textit{eigenvalue problem} on $\omega^2$. The ultimate goal of this work is to compute the negative eigenvalues since they give the growth rate of the gravitational instability.

\section{Reformulating the governing equations}
\label{section:Reformulation}

In this section, we first rewrite the governing system of equations \eqref{EigenvalueProblem} in a far more convenient form, namely \eqref{EqnIntheta} below. Then we introduce a notation that will be very handy for the calculations, namely matrix $\mathsfbi{H}$ in \eqref{MatrixH}. Finally, from the boundary conditions we derive a scalar equation, namely \eqref{Quantization}, whose solutions are the eigenvalues.

\subsection{Reformulated equations of motion}

First of all, let us work with dimensionless variables. From the homogeneous case \citep[e.g.][]{Thompson}, it is well known that a key length scale in gravitational instability is given by the critical Jeans wavenumber
\begin{equation}
k_{J}(x) \equiv \sqrt{\frac{4 \pi G \rho_0(x)}{c_a^2(x)}}
\end{equation}
which marks the balance between pressure and gravity. In this definition $k_J$ is position dependent since the equilibrium density profile $\rho_0$ is not assumed homogeneous here. We use the central value (subscripts $c$) of this quantity, noted as
\begin{equation}
k_{Jc} \equiv k_{J}(x=0) = \sqrt{\frac{4 \pi G}{\kappa \gamma} \rho_c^{2-\gamma}}
\end{equation}
where we used definition \eqref{ca2} of $c_a^2$ in the last equality, together with the central value of the density $\rho_c \equiv \rho_0(x=0)$, to define the following dimensionless parameters
\begin{equation}
\begin{array}{l}
\ \\
\left\{
\begin{array}{llll}
\bar{x} \equiv k_{Jc} x & \rho \equiv \frac{\rho_0}{\rho_c} & \bar{\omega}^2 \equiv \frac{\omega^2}{4 \pi G \rho_c} & \bar{k}_y \equiv \frac{k_y}{k_{Jc}} \\
\psi \equiv \rho k_{Jc} \hat{\xi}_x & \mathcal{R} \equiv \rho^{\gamma-2} \frac{\hat{\rho}_1}{\rho_c} & \mathcal{G}_x \equiv \frac{k_{Jc}}{4 \pi G \rho_c} \hat{g}_{1x} & \mathcal{G}_y \equiv \frac{k_{Jc}}{4 \pi G \rho_c} i \bar{k}_y \hat{g}_{1y}\\
\end{array}
\right.
\end{array} .
\label{DimensionlessVariables}
\end{equation}
All these definitions are straightforward, except that of the variable $\mathcal{R}$ (this letter referring to `rho', from $\rho_1$).  We choose to work with $\mathcal{R} \equiv \rho_0^{\gamma-2} \hat{\rho}_1 / \rho_c^{\gamma-1}$ rather than with the more natural $\hat{\rho}_1 / \rho_c$, because we notice that, using the definition \eqref{ca2} of $c_a^2$ and the hydrostatic equilibrium \eqref{HydroEquilibrium} we have the relation
\begin{equation}
c_a^2 \hat{\rho}_1' + \left(-\hat{g}_0 + (c_a^2)'\right) \hat{\rho}_1 = \kappa \gamma \rho_0 \left(\rho_0^{\gamma-2} \hat{\rho}_1\right)',
\label{trick}
\end{equation}
which turns the second equation in \eqref{EigenvalueProblem} into a more compact form, such that we may rewrite the whole system of equations into a matrix form that is by blocks. Indeed, with variables \eqref{DimensionlessVariables}, the system \eqref{EigenvalueProblem} reads
\begin{equation}
\frac{\dd}{\dd\bar{x}}
\left(
\begin{array}{c}
\psi\\
\mathcal{G}_x\\
\mathcal{R}\\
\mathcal{G}_y
\end{array}
\right)
=
\left(
\begin{array}{cccc}
0 & 0 & \frac{\bar{k}_y^2}{\bar{\omega}^2} \rho(\bar{x}) - \rho(\bar{x})^{2-\gamma} & \frac{\rho(\bar{x})}{\bar{\omega}^2} \\
0 & 0 & - \rho(\bar{x})^{2-\gamma} & -1 \\
\frac{\bar{\omega}^2}{\rho(\bar{x})} & 1 & 0 & 0 \\
0 & -\bar{k}_y^2 & 0 & 0 \\
\end{array}
\right)
\left(
\begin{array}{c}
\psi\\
\mathcal{G}_x\\
\mathcal{R}\\
\mathcal{G}_y
\end{array}
\right)
\label{EqnInxbar}
\end{equation}
where we plugged the third equation of \eqref{EigenvalueProblem} into its first equation. This is already better than the initial formulation, but it is still very complicated, because we do not know what the coefficients in this matrix equation look like explicitly. Indeed, to do so we would need to solve for $\rho(\bar{x})$, i.e. integrate the Lane-Emden equation \eqref{LaneEmden}, but as stated in section \ref{section:EquilibriumState}, this has simple analytic solutions only for the special cases $\gamma = \frac{2}{3}, 1, 2$ and $\infty$. And even in these cases in which the equilibrium state is simple, the equation above, governing the evolution of the perturbations, is not simple. For example with $\gamma=2$ we have $\rho(\bar{x}) = \cos(\bar{x})$, and $\gamma=1$ we have $\rho(\bar{x}) = \cosh(\bar{x}/\sqrt{2})^{-2}$, so that the differential equation \eqref{EqnInxbar} is deceivingly complicated, especially for arbitrary $\gamma$.

To improve this, let us do the change of variables (illustrated in figure~\ref{fig:SchemaPerturbedProfile})
\begin{equation}
\rho = \frac{\rho_0(\bar{x})}{\rho_c},
\label{ChangeVariable_xRho}
\end{equation}
which is possible because the equilibrium density profile $\rho_0$ is monotonic so the change $\bar{x} \leftrightarrow \rho$ is bijective. Interestingly, this change of variables is implicit since we do not need to know the functional form of $\rho_0(\bar{x})$. Indeed, using the chain rule, $\rho'$ is simple to compute: With the dimensionless variables \eqref{DimensionlessVariables}, the Lane-Emden relation \eqref{LaneEmden} becomes
\begin{equation}
(\rho^{\gamma-1})''+(\gamma-1)\rho=0,
\label{LaneEmden_Dimensionless}
\end{equation}
and multiplying by $(\rho^{\gamma-1})'$, it can be rewritten as
\begin{equation}
\frac{\dd}{\dd\bar{x}} \left[\frac{1}{2}\left(\left(\rho^{\gamma-1}\right)'\right)^2+\frac{(\gamma-1)^2}{\gamma} \rho^\gamma\right] = 0.
\end{equation}
We integrate this, determining the constant of integration by considering profiles that are flat at the centre, i.e.\ $\rho'(\bar{x}=0)=0$, which is a natural choice. Then we get
\begin{equation}
\frac{\dd \rho}{\dd\bar{x}} = - \sqrt{\frac{2}{\gamma}} \rho^{2-\gamma} \sqrt{1-\rho^{\gamma}}
\label{rho0prime}
\end{equation}
where we took the solution with negative sign because the profile is decreasing with $\bar{x}$. Equation \eqref{EqnInxbar} can thus be rewritten as
\begin{equation}
\frac{\dd}{\dd\rho}
\left(
\begin{array}{c}
\psi\\
\mathcal{G}_x\\
\mathcal{R}\\
\mathcal{G}_y
\end{array}
\right)
=
- \sqrt{\frac{\gamma}{2}} \frac{\rho^{\gamma-2}}{\sqrt{1-\rho^\gamma}}
\left(
\begin{array}{cccc}
0 & 0 & \frac{\bar{k}_y^2}{\bar{\omega}^2} \rho - \rho^{2-\gamma} & \frac{\rho}{\bar{\omega}^2} \\
0 & 0 & - \rho^{2-\gamma} & -1 \\
\frac{\bar{\omega}^2}{\rho} & 1 & 0 & 0 \\
0 & -\bar{k}_y^2 & 0 & 0 \\
\end{array}
\right)
\left(
\begin{array}{c}
\psi\\
\mathcal{G}_x\\
\mathcal{R}\\
\mathcal{G}_y
\end{array}
\right).
\label{EqnInrho}
\end{equation}
Comparing \eqref{EqnInxbar} and \eqref{EqnInrho} the modification may not look spectacular, but it is in fact a huge progress: Equation \eqref{EqnInxbar} is very complicated because it has complicated coefficients that are not even explicit (in its variable $\bar{x}$), while \eqref{EqnInrho} contains simple coefficients that are explicit (in its variable $\rho$).

Finally, the occurrence of the square root $\sqrt{1-\rho^\gamma}$ in \eqref{EqnInrho} calls for another change of variables (illustrated in figure~\ref{fig:SchemaPerturbedProfile}), namely
\begin{equation}
\sin \theta = \rho^{\gamma/2},
\label{ChangeVariable_RhoTheta}
\end{equation}
which makes sense since $\rho \in [0,1]$ by definition, and as we shall see below, this will indeed facilitate greatly our calculations. Note that, in order to lighten expressions, we will often use the shorthand notations
\begin{equation}
\left\{
\begin{array}{l}
s \equiv \sin \theta\\
c \equiv \cos \theta
\end{array}.
\right.
\end{equation}
Therefore, in this paper we will work with the following form of system \eqref{EigenvalueProblem}
\begin{equation}
\frac{\dd}{\dd\theta}
\left(
\begin{array}{c}
\psi\\
\mathcal{G}_x\\
\mathcal{R}\\
\mathcal{G}_y
\end{array}
\right)
=
- \sqrt{\frac{2}{\gamma}}
\left(
\begin{array}{cccc}
0 & 0 & \frac{\bar{k}_y^2}{\bar{\omega}^2} s - s^{\frac{2}{\gamma}-1} & \frac{s}{\bar{\omega}^2} \\
0 & 0 & - s^{\frac{2}{\gamma}-1} & -s^{1-\frac{2}{\gamma}} \\
\bar{\omega}^2 s^{1-\frac{4}{\gamma}} & s^{1-\frac{2}{\gamma}} & 0 & 0 \\
0 & -\bar{k}_y^2 s^{1-\frac{2}{\gamma}} & 0 & 0 \\
\end{array}
\right)
\left(
\begin{array}{c}
\psi\\
\mathcal{G}_x\\
\mathcal{R}\\
\mathcal{G}_y
\end{array}
\right).
\label{EqnIntheta}
\end{equation}

Let us now define the matrix $\mathsfbi{H}$ such that
\begin{equation}
\left(
\begin{array}{r}
\mathcal{R}(\theta)\\
\mathcal{G}_y(\theta)\\
\end{array}
\right)
 = \mathsfbi{H}(\theta)
\left(
\begin{array}{r}
\mathcal{R}(\tfrac{\pi}{2})\\
\mathcal{G}_y(\tfrac{\pi}{2})
\end{array}
\right)
\label{PGy_AvecH}
\end{equation}
i.e. $\mathsfbi{H}$ describes the variation with position $\theta$ of $(\mathcal{R},\mathcal{G}_y)$ with respect to their value at the centre ($\theta = \pi/2$). We will use subscripts $a=1,2,3$ and $4$ to denote its coefficients, namely
\begin{equation}
\mathsfbi{H} =
\left(
\begin{array}{cc}
h_1 & h_2\\
h_3 & h_4
\end{array}
\right)
\hspace{1cm} \mathrm{where} \ h_a\equiv h_a(\theta;\bar{\omega}^2,\bar{k}_y^2).
\label{MatrixH}
\end{equation}
The matrix $\mathsfbi{H}$ characterizes completely the eigenfunctions since, by the definition \eqref{PGy_AvecH}, $\mathsfbi{H}(\theta)$ gives the value of $\mathcal{R}$ and $\mathcal{G}_y$ at a given~$\theta$, and the two other eigenfunctions can be deduced, either in integral form by integrating the  first two rows of \eqref{EqnIntheta}, namely
\begin{equation}
\left(
\begin{array}{c}
\psi\\
\mathcal{G}_x\\
\end{array}
\right)
 = - \sqrt{\tfrac{2}{\gamma}} \int_{\frac{\pi}{2}}^\theta \dd\theta_1
\left(
\begin{array}{cc}
\frac{\bar{k}_y^2}{\bar{\omega}^2} s - s^{\frac{2}{\gamma}-1} & \frac{s}{\bar{\omega}^2} \\
- s^{\frac{2}{\gamma}-1} & -s^{1-\frac{2}{\gamma}}
\end{array}
\right)
 \mathsfbi{H}(\theta_1)
\left(
\begin{array}{r}
\mathcal{R}(\tfrac{\pi}{2})\\
\mathcal{G}_y(\tfrac{\pi}{2})
\end{array}
\right)
\label{PsiGx_WithIntegralOfA}
\end{equation}
(where we have used the symmetry conditions \eqref{SymmetryConditions_newVariables} below) or expressed with a derivative by inverting the  last two rows of \eqref{EqnIntheta}, namely
\begin{equation}
\left(
\begin{array}{c}
\psi\\
\mathcal{G}_x\\
\end{array}
\right)
 = 
\sqrt{\frac{\gamma}{2}} \frac{s^{\frac{4}{\gamma}-1}}{\bar{\omega}^2 \bar{k}_y^2}
\left(
\begin{array}{cc}
-\bar{k}_y^2 & - 1 \\
0 & \bar{\omega}^2 s^{-\frac{2}{\gamma}} \\
\end{array}
\right)
\mathsfbi{H}'(\theta)
\left(
\begin{array}{r}
\mathcal{R}(\tfrac{\pi}{2})\\
\mathcal{G}_y(\tfrac{\pi}{2})
\end{array}
\right).
\label{PsiGx_WithInverseB}
\end{equation}

\subsection{Reformulated boundary conditions and symmetry conditions}
\label{section:BC}

Let us start with two preliminary remarks. First, from now on, subscripts $b$ will indicate values taken at the boundary, notably $\bar{x}_b, \rho_b$ and $\theta_b$. Secondly, note that an interesting feature of the change of variable \eqref{ChangeVariable_xRho} is that thanks to it, in this work we will derive the eigenvalues without integrating the Lane-Emden equation (which is possible only in a few cases), i.e. without having to obtain the equilibrium density as a function of position $\rho_0(x)$. What matters is not where the BC is imposed (position $\bar{x}_b$), but the density $\rho_b$ at the position where it is imposed. This is why in the end, the explicit formul{\ae} \eqref{omega02_coeffs_explicit} are expressed in terms of $\rho_b$.
Hence, given the importance of $\rho_b$, let us see explicitly how it is related to the external pressure applied on the structure.
With the equation of state \eqref{p0_polytropic} and from the definition \eqref{Def:xb} of $x_b$ we have
\begin{equation}
\rho_b = \left(\frac{p_\mathrm{ext}}{p_c}\right)^{1/\gamma}
\label{Link_rhob_pext}
\end{equation}
where $p_c \equiv p_0(x=0) = \kappa \rho_c^\gamma$ is the central value of the equilibrium pressure.

We now update our  formulation of section \ref{section:BC_firstFormulation} using the new variables \eqref{DimensionlessVariables}. As far as symmetry conditions are concerned, it is clear that \eqref{SymmetryConditions} becomes
\begin{equation}
\left\{
\begin{array}{l}
\psi\left(\tfrac{\pi}{2}\right) = 0\\
\mathcal{G}_x\left(\tfrac{\pi}{2}\right) = 0.
\end{array}
\right.
\label{SymmetryConditions_newVariables}
\end{equation}
Then, the BC \eqref{BC1_OldVariables} reads
\begin{equation}
\psi(\theta_b) - \mathcal{G}_x(\theta_b) + \bar{k}_y^{-1} \mathcal{G}_y(\theta_b) = 0.
\label{BC1}
\end{equation}
We also combine  \eqref{BC2_Rigid_OldVariables} and \eqref{BC2_Free_OldVariables}  into a single expression,
\begin{equation}
\psi(\theta_b) + \Delta_\mathrm{BC} \ \mathcal{R}(\theta_b) = 0
\label{BC2}
\end{equation}
where
\begin{equation}
\Delta_\mathrm{BC} \equiv \delta_{\mathrm{BC}} \frac{4 \pi G \rho_c}{k_{Jc}} \frac{\rho_b}{g_0(\bar{x}_b)}
\end{equation}
and we defined the following parameter
\begin{equation}
\delta_{\mathrm{BC}} \equiv
\left\{
\begin{array}{ll}
\displaystyle
1 & \mathrm{for \ Free \ BC}\\
\displaystyle
0 & \mathrm{for \ Rigid \ BC}
\end{array}
\right.
\label{delta_BC}
\end{equation}
which enables us to switch between the two types of BCs and spares us the pain of writing separate equations. With the definition of the sound speed \eqref{ca2} and the hydrostatic equilibrium \eqref{HydroEquilibrium} we may rewrite $\Delta_\mathrm{BC}$ using the $\rho$ and $\theta$ variables, as
\begin{equation}
\Delta_\mathrm{BC} = - \delta_{\mathrm{BC}} \sqrt{\frac{\gamma}{2}} \frac{\rho_b}{\sqrt{1-\rho_b^\gamma}} = - \delta_{\mathrm{BC}} \sqrt{\frac{\gamma}{2}} \frac{(\sin \theta_b)^{2/\gamma}}{\cos \theta_b}.
\label{Zb}
\end{equation}

\subsection{Eigenvalue equation}

From the BC above, let us derive a single scalar equation which constrains the eigenvalue parameter $\bar{\omega}^2$. We rewrite the boundary conditions \eqref{BC1} and \eqref{BC2} in matrix form as
\begin{equation}
\left(
\begin{array}{cc}
1 & -1\\
1 & 0
\end{array}
\right)
\left(
\begin{array}{c}
\psi(\theta_b)\\
\mathcal{G}_x(\theta_b)\\
\end{array}
\right)
+
\left(
\begin{array}{cc}
0 & 1/\bar{k}_y\\
\Delta_\mathrm{BC} & 0
\end{array}
\right)
\left(
\begin{array}{c}
\mathcal{R}(\theta_b)\\
\mathcal{G}_y(\theta_b)\\
\end{array}
\right)
=
\left(
\begin{array}{c}
0\\
0
\end{array}
\right),
\label{BC_intermediateForm}
\end{equation}
and using \eqref{PGy_AvecH} and \eqref{PsiGx_WithInverseB}, we rewrite this with the coefficients of matrix $\mathsfbi{H}$ and their derivatives, namely
\begin{equation}
\left(
\begin{array}{cc}
q_1(\theta_b) & q_2(\theta_b)\\
q_3(\theta_b) & q_4(\theta_b)
\end{array}
\right)
\left(
\begin{array}{c}
\mathcal{R}\left(\tfrac{\pi}{2}\right)\\
\mathcal{G}_y\left(\tfrac{\pi}{2}\right)
\end{array}
\right)
=
\left(
\begin{array}{c}
0\\
0
\end{array}
\right)
\label{BC_MatrixForm}
\end{equation}
where (all quantities below are evaluated at $\theta = \theta_b$)
\begin{equation}
\left\{
\begin{array}{l}
q_1 = \bar{k}_y^2 \ h'_1 + [1 + s^{-\frac{2}{\gamma}} \bar{\omega}^2] \ h'_3 - \! \sqrt{\tfrac{2}{\gamma}} s^{1-\frac{4}{\gamma}} \bar{\omega}^2 \bar{k}_y \ h_3\\
q_2 =  \bar{k}_y^2 \ h'_2 + [1 + s^{-\frac{2}{\gamma}} \bar{\omega}^2] \ h'_4 - \! \sqrt{\tfrac{2}{\gamma}} s^{1-\frac{4}{\gamma}} \bar{\omega}^2 \bar{k}_y \ h_4\\
q_3 = \bar{k}_y^2 \ h'_1 + h'_3 - \! \sqrt{\tfrac{2}{\gamma}} s^{1-\frac{4}{\gamma}} \bar{\omega}^2  \Delta_\mathrm{BC} \bar{k}_y^2 \ h_1\\
q_4 = \bar{k}_y^2 \ h'_2 + h'_4 - \! \sqrt{\tfrac{2}{\gamma}} s^{1-\frac{4}{\gamma}} \bar{\omega}^2  \Delta_\mathrm{BC} \bar{k}_y^2 \ h_2
\end{array} .
\right.
\label{Def:qa}
\end{equation}
The matrix in \eqref{BC_MatrixForm} must be non invertible for non trivial solutions to exist: its determinant must vanish,
\begin{equation}
q_1 q_4 - q_3 q_2 = 0.
\label{Quantization}
\end{equation}
This relation is the starting point of our analysis. It is an equation on $\bar{\omega}^2$, and its solutions are the eigenvalues that we are looking for. For a given $\bar{k}_y$ these solutions constitute the discrete spectrum we are looking for, and we will refer to it as the `eigenvalue equation'. In the remainder of this paper, we detail how we solve equation \eqref{Quantization} and obtain the growth rate of gravitational instability.

\section{Growth rate}
\label{section:Method}

We now present the method we introduce to find the eigenvalues $\bar{\omega}^2$. To the best of our knowledge, it has not been introduced in the literature yet.

Relation \eqref{Quantization} is the equation on $\bar{\omega}^2$ that we want to solve. In that form however, part of its dependence on  $\bar{\omega}^2$  is implicit. Thus, let us first state it explicitly. As we see from their definitions \eqref{Def:qa}, the $q_a$'s are linear combinations of the $h_a$'s and their derivatives. The $h_a$'s are unknown at this stage, but we know that they are parameterised by $\bar{\omega}^2$ and $\bar{k}_y^2$, i.e.\ $h_a = h_a(\theta;\bar{\omega}^2,\bar{k}_y^2)$, as stated in \eqref{MatrixH}. Let us expand the $h_a$'s in powers of $\bar{\omega}^2$, using the notation
\begin{equation}
h_a = \sum_{i=0}^\infty h_a^i \ (\bar{\omega}^2)^i \hspace{1cm} \mathrm{where} \ h_a^i\equiv h_a^i(\theta;\bar{k}_y^2).
\label{Def:hia}
\end{equation}
Doing so, the $q_a$'s are in the form
\begin{equation}
q_a = \sum_{i=0}^\infty q_a^i \ (\bar{\omega}^2)^i,
\end{equation}
where it is straightforward to obtain the coefficients $q_a^i$ from \eqref{Def:qa} by gathering terms in powers of $\bar{\omega}^2$. Relation \eqref{Quantization} then becomes
\begin{equation}
\sum_{k=0}^\infty a_k \ (\bar{\omega}^2)^k = 0
\label{InfinitePolynomial}
\end{equation}
where, using the Cauchy formula for the product of two infinite series,
\begin{equation}
a_k = \sum_{\ell=0}^k (q_1^\ell q_4^{k-\ell} - q_3^\ell q_2^{k-\ell}).
\label{ak}
\end{equation}
Thus we wrote relation \eqref{Quantization} as a power series in $\bar{\omega}^2$. The point is that the zeros of this power series are the eigenvalues we are looking for.  They constitute a discrete set of values, the discrete spectrum of the eigenvalue problem \eqref{EigenvalueProblem}. In a way, by analogy with Quantum Mechanics, equation \eqref{Quantization} [or equivalently \eqref{InfinitePolynomial}] could be called a `quantization relation'. The power series is a priori of infinite degree, which means that there may be an infinite number of eigenvalues\footnote{We expect this since, as it is well known, already in the Cowling approximation (see below) we end up with a Sturm-Liouville problem \citep[e.g.][]{Cowling41}, in which case the existence of an infinite number of well ordered eigenfrequencies can be shown rigorously.}.
Let us denote these zeros by $\bar{\omega}_n^2$, with $n \geq 0$. As many previous studies have already shown,
there is a lowest eigenfrequency $\bar{\omega}_0^2$, called the fundamental, which becomes negative for some values of $\bar{k}_y$, and which gives rise to the gravitational instability. However, previous authors did not provide ways to derive general analytic expressions for the fundamental $\bar{\omega}_0^2$ as a function of $\bar{k}_y$, which is our aim here.

Now, at first sight relation \eqref{InfinitePolynomial} may not look very encouraging, because finding the roots of a polynomial of order greater than 3 is extremely difficult, if not impossible. However, we can go around this difficulty as follows. Consider a polynomial in $\bar{\omega}^2$ of finite order $N$, with coefficients $a_n$ and roots $\bar{\omega}_n^2$. We have
\begin{equation}
\sum_{n=0}^N a_n (\bar{\omega}^2)^n = a_N \prod_{n=0}^{N-1} (\bar{\omega}^2 - \bar{\omega}_n^2),
\end{equation}
so that, identifying the $(\bar{\omega}^2)^0$ and $(\bar{\omega}^2)^1$ terms from both sides, we get respectively
\begin{equation}
a_0 = (-1)^N a_N \prod_{n=0}^{N-1} \bar{\omega}_n^2
\label{a0}
\end{equation}
and
\begin{equation}
a_1 = (-1)^{N-1} a_N \sum_{k=0}^{N-1} \prod_{\substack{n=0 \\ n \neq k}}^{N-1} \bar{\omega}_n^2.
\label{a1}
\end{equation}
These are simply the two first `root-coefficient relations', also known as Vieta's formulae.
Dividing \eqref{a1} by \eqref{a0} and taking the limit $N \rightarrow \infty$, we can express the fundamental eigenfrequency in terms of the higher order modes, namely
\begin{equation}
\bar{\omega}^2_0 = - \left[\frac{a_1}{a_0} + \sum_{n=1}^{\infty} \frac{1}{\bar{\omega}^2_n}\right]^{-1}.
\label{LinkBetweenEigenfrequencies}
\end{equation}
This is the key formula in our derivation. It is interesting per se, since it is a link between the fundamental mode $\bar{\omega}^2_0$, which is unstable (in some ranges of $\bar{k}_y$), and the higher order modes $\bar{\omega}^2_{n \geq 1}$, which are stable (acoustic oscillations). Therefore, interestingly, in this paper we are going to compute with formula \eqref{LinkBetweenEigenfrequencies} the unstable part of the spectrum using its stable part, the latter being by far much easier to compute. In addition, by doing so, we will manage to decompose the initial full fourth-order problem
into a sequence of second-order problems that can be solved separately. More precisely, we are going to compute the right hand side of formula \eqref{LinkBetweenEigenfrequencies} in the three following steps.

\subsection*{Step I : Computing $\sum_{n=1}^{\infty} 1/\bar{\omega}^2_n$}

Let us first focus on the series appearing in formula \eqref{LinkBetweenEigenfrequencies}. An outline is given here, and details are provided in Appendix \ref{section:StepI}. Interestingly, at first sight relation \eqref{LinkBetweenEigenfrequencies} does not look like a strategic path to take at all. Indeed, it seems that in order to solve \textit{one} fourth order problem (computing $\bar{\omega}^2_0$), we will have to solve \textit{an infinite} number of fourth order problems (computing the $\bar{\omega}^2_{n \geq 1}$'s) and in addition we will need to compute a series, which could in principle be extremely difficult too. However, the key point is to notice that in fact, computing the $\bar{\omega}^2_{n \geq 1}$'s constitutes a \textit{much} easier problem than the initial one because, as in stellar oscillation theory, we may compute these high order modes\footnote{We borrow here the terminology from  Asteroseismology.} with extremely high accuracy in the so-called Cowling approximation \citep{Cowling41}. It consists in neglecting the perturbation of the gravitational field $\vec{g}_1$ in the linearised momentum conservation \eqref{LinearizedMomentumConservation}, in which case the problem becomes only second order, instead of fourth. As a reminder of this, we will use the notation
\begin{equation}
\mathscr{C} \equiv \sum_{n=1}^{\infty} \frac{1}{\bar{\omega}^2_n}
\label{Def:C}
\end{equation}
where the letter $\mathscr{C}$ stands for `Cowling'.

Though by far much simpler than in the full case, the equations in the Cowling approximation remain difficult to solve exactly. The second key point now is that we know that the $\bar{\omega}^2_{n \geq 1}$'s are large, with even $\bar{\omega}_n^2 \rightarrow \infty$ as $n \rightarrow \infty$, so that when we want to solve our eigenvalue problem looking for the high order modes, we know that  the parameter $\bar{\omega}$ will be large. Therefore, we may use the Wentzel - Kramers - Brillouin (WKB) method treating $1/\bar{\omega}$ as a small parameter.

\begin{figure}
\centering
\includegraphics[scale=0.6]{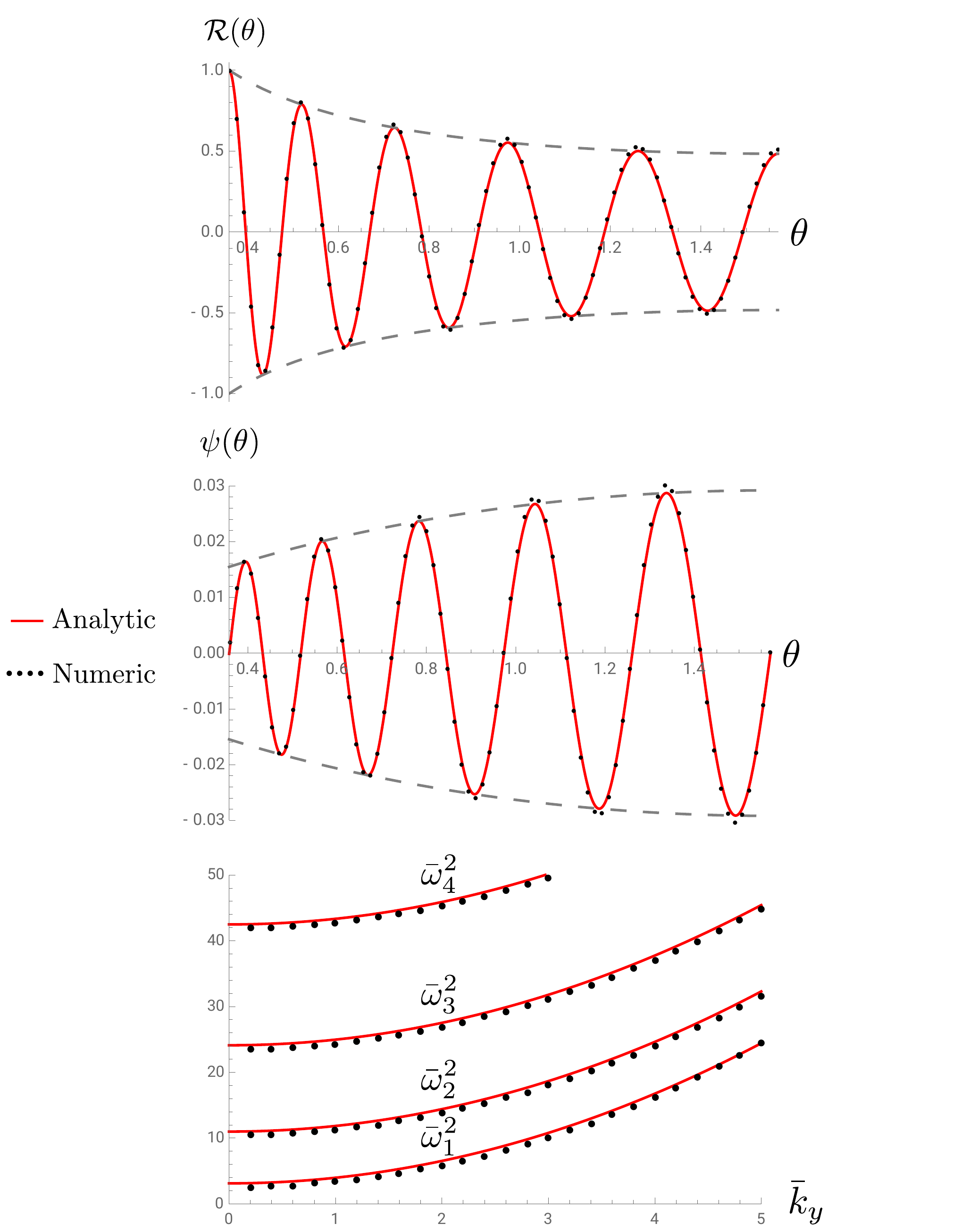}
\caption{In Appendix~\ref{section:StepI} we compute the stable part of the spectrum (i.e.\ harmonics $n \geq 1$). Here we compare our analytic expressions (red lines) with the numerical resolution (black dots) for the eigenfunction $\mathcal{R}(\theta)$ in the top panel, $\psi(\theta)$ at the centre, and for the eigenvalues $\bar{\omega}_n^2$ given by \eqref{Spectrum_Cowling} at the bottom (shown are $n=1,2,3$ and $4$). In the top and middle panels, the $x$-axis ranges from $\theta_b$ (the boundary) to $\pi/2$ (centre of the slab). In this example we chose the Rigid BC (hence the fact that the curve for $\psi(\theta)$ reaches zero on the left and right of its plot), with values $\gamma = 1.3$, $\rho_b = 0.2$, $n=10$ and $\bar{k}_y = 2.5$. Normalization has been chosen such that $\mathcal{R}(\theta_b) = 1$. Gray dashed lines indicate the envelopes, such as the one appearing in front of the cosine in expression \eqref{WKB_solution}. Note that the numerical resolution for the eigenfunctions here is done in the Cowling approximation, to show that the WKB resolution gives very satisfying results (such an agreement between the analytic and numerical solutions happens for a very wide range of parameters). However, in the bottom panel, the analytic result is compared to the \textit{full} numerical result, i.e. not in the Cowling approximation. Therefore, the bottom plot shows that the approximate spectrum \eqref{Spectrum_Cowling} fits the full numerical result very well, which validates our procedure of combining the Cowling approximation with the WKB method, and makes expression \eqref{CowlingTerm} for $\mathscr{C}$ very efficient.}
\label{Cowling_Eigenfunctions_And_Eigenvalues}
\end{figure}

With the above two ideas (Cowling \& WKB) we derive in Appendix~\ref{section:StepI} very precise formulas for the eigenfunctions $\mathcal{R}$ and $\psi$   (the only relevant ones in the Cowling approximation) of equation (\ref{EqnInxbar}), and the eigenvalues $\bar{\omega}^2_{n \geq 1}$'s. In the top and middle panels of figure~\ref{Cowling_Eigenfunctions_And_Eigenvalues}, we show an example of a plot of $\mathcal{R}$ and $\psi$ using these analytic expressions, compared to the numerical resolution. For a very wide range of the parameters $(\gamma, \rho_b, n, \bar{k}_y)$ our WKB resolution gives excellent results. In the bottom panel, we show the spectrum $\bar{\omega}^2_{n \geq 1}$ that we obtain with formula \eqref{Spectrum_Cowling} and compare it to the spectrum solved numerically using a shooting method. As we can see again, both agree extremely well. Let us stress that in the latter plot, the numerical solution corresponds to the full resolution of the initial eigenvalue problem \eqref{EigenvalueProblem}, without the Cowling approximation. Therefore, this panel shows that it is not only the WKB resolution itself that works very well (as illustrated in the top and middle panels of figure~\ref{Cowling_Eigenfunctions_And_Eigenvalues} for the eigenfunctions) but also the Cowling approximation, since the eigenvalues \eqref{Spectrum_Cowling} were obtained by combining both.

Then finally, in Appendix~\ref{section:StepI}, we compute the series \eqref{Def:C} 
with the pleasant satisfaction of being able to express it in a simple closed form,
\begin{equation}
\mathscr{C} = \frac{1}{2 \ \! A \ \! b_\mathrm{BC}^2} \left[\pi b_\mathrm{BC} \tanh\left(\pi b_\mathrm{BC}\right)^{2 \delta_\mathrm{BC} -1} + \delta_\mathrm{BC} -1 \right]
\label{CowlingTerm}
\end{equation}
where
\begin{equation}
\left\{
\begin{array}{l}
\displaystyle
b_\mathrm{BC} = \sqrt{\frac{B \ \bar{k}_y^2 + C + D_\mathrm{BC}}{A}}\\[1.em]
\displaystyle
A = \frac{\gamma \pi^2}{2 I_{-\frac{1}{\gamma}}(\theta_b)}\\[1.5em]
\displaystyle
B = I_{2-\tfrac{3}{\gamma}}(\theta_b)\\[1.em]
\displaystyle
C = \frac{(\gamma -3)}{8 \gamma} \left[(5-\gamma) \ I_{\tfrac{1}{\gamma}}(\theta_b) + (3 \gamma -5) \ I_{\tfrac{1}{\gamma}-2}(\theta_b)\right]\\[1.em]
\displaystyle
D_\mathrm{BC} = -2 \sin(\theta_b)^{\frac{1}{\gamma}-1} \cos(\theta_b) \left(1 + (\delta_{\mathrm{BC}}-1) \frac{1+\gamma}{4}\right),
\end{array}
\right.
\label{Coeffs_A_B_C_D}
\end{equation}
the subscripts `BC' denoting quantities that depend on the type of BCs ($\delta_{\mathrm{BC}} = 1$ for Free BC, $0$ for Rigid BC), and we use the notation
\begin{equation}
I_p(\theta) \equiv \int_{\frac{\pi}{2}}^{\theta} (\sin\theta)^p \dd\theta = 2^p \left[B_{\frac{1-\cos \theta}{2}}\left(\tfrac{p+1}{2},\tfrac{p+1}{2}\right) - B_{\frac{1}{2}}\left(\tfrac{p+1}{2},\tfrac{p+1}{2}\right)\right],
\label{Def:Ip}
\end{equation}
where the second equality comes from the definition of the incomplete Beta function.

\subsection*{Step II : Computing $a_1/a_0$}

Let us now focus on the other term, $a_1/a_0$, contributing to $\bar \omega_0^2 $ in formula \eqref{LinkBetweenEigenfrequencies}, details being provided in Appendix \ref{section:StepII}. First of all, it is interesting to note that thanks to \eqref{LinkBetweenEigenfrequencies}, we need to compute only two of the coefficients $a_k$ of the power series \eqref{InfinitePolynomial}. What is more, these first two are also the simplest. Then, from  definition \eqref{ak}, it turns out that the only things needed are the functions $h_a^0$ and $h_a^1$ of expansion \eqref{Def:hia}. That is, out of an a priori infinite number of functions, we need only eight (since $a = 1,2,3$ and $4$). Now, as detailed in Appendix \ref{section:StepII}, in order to obtain $h_a^0$ and $h_a^1$, we first derive the differential equations they satisfy from the equation of motion \eqref{EqnIntheta}. Then, recalling that for gravitational instability only large scale perturbations are relevant (i.e. small $\bar{k}_y$, otherwise pressure balances gravity), we look for the solutions by expanding $h_a^0$ and $h_a^1$ in powers of $\bar{k}_y$ (cf. definition \eqref{Def:hiaj_a0m_a1m} below). The practical advantage of this expansion is that, while $h_a^0$ and $h_a^1$ satisfy fourth order equations, the problem is actually reduced to a(n infinite) sequence of \textit{second order} problems that are \textit{fully-solvable}, namely equation \eqref{eqn_hiaj} in the Appendix. Therefore, we compute $h_a^0$ and $h_a^1$ \textit{exactly}, but order by order in $\bar{k}_y$. For this reason, in the end we obtain $a_0$, $a_1$, and consequently $\bar{\omega}_0^2$, as power series in $\bar{k}_y$. We show the general procedure in Appendix \ref{section:StepII}, but let us give here only the intermediary expressions necessary to derive $\bar{\omega}_0^2$ up to the second order in $\bar{k}_y$. Denoting the expansions in $\bar{k}_y$ by
\begin{equation}
h_a^i = \sum_{j=0}^\infty h_{a,j}^i \left(\bar{k}_y^2\right)^j \hspace{0.5cm} , \hspace{0.5cm} a_0 =  \sum_{m=0}^\infty a_{0,m} (\bar{k}_y)^m \hspace{0.5cm} \mathrm{and} \hspace{0.5cm} a_1 =  \sum_{m=0}^\infty a_{1,m} (\bar{k}_y)^m,
\label{Def:hiaj_a0m_a1m}
\end{equation}
we obtain that the useful $h_{a,j}^i$'s are given by
\begin{equation}
\left\{
\begin{array}{l}
h_{1,0}^0(\theta) = \left[Q_{\frac{1}{\gamma}}^{1-\frac{1}{\gamma}}(0)\right]^{-1} \ y_Q(\theta)\\
h_{1,0}^1(\theta) = \tfrac{2}{\gamma} \Gamma\left(\tfrac{2}{\gamma}\right) \left[Q_{\frac{1}{\gamma}}^{1-\frac{1}{\gamma}}(0)\right]^{-1} \ \left[ y_Q(\theta) \! \int_{\frac{\pi}{2}}^\theta \ \left(\frac{c}{s} y_Q'+y_Q\right) \ \frac{y_P}{s} \ \dd\theta_1 - \ y_P(\theta) \! \int_{\frac{\pi}{2}}^\theta \ \left(\frac{c}{s} y_Q'+y_Q\right) \ \frac{y_Q}{s} \ \dd\theta_1 \right]
\end{array}
\right.
\label{h010_and_h110}
\end{equation}
where $\Gamma$ is the Gamma function and we defined
\begin{equation}
\left\{
\begin{array}{l}
y_P(\theta) \equiv (\sin \theta)^{1-\frac{1}{\gamma}} P_{\frac{1}{\gamma}}^{1-\frac{1}{\gamma}}(\cos \theta)\\
y_Q(\theta) \equiv (\sin \theta)^{1-\frac{1}{\gamma}} Q_{\frac{1}{\gamma}}^{1-\frac{1}{\gamma}}(\cos \theta)\\
\end{array}
\right.
\label{Def:yP_yQ}
\end{equation}
with $P_\lambda^\mu$ and $Q_\lambda^\mu$ the associated Legendre functions of the first and second kind. The useful $a_{n,m}$'s are given by
\begin{equation}
\left\{
\begin{array}{l}
a_{0,1} = \frac{2}{\gamma} c s^{\frac{2}{\gamma}} \left((h_{1,0}^0)' + \delta_{\mathrm{BC}} \frac{s}{c} h_{1,0}^0\right)\\
a_{1,0} = \sqrt{\frac{2}{\gamma}} s^{\frac{2}{\gamma}} \left((h_{1,0}^0)' + \delta_{\mathrm{BC}} \frac{s}{c} h_{1,0}^0\right)
\end{array}
\right.
\label{a01_a10}
\end{equation}
for the linear expansion of $\bar \omega_0^2$ (cf. \eqref{omega02_coeffs_a} below), while those for the quadratic expansion (cf. \eqref{omega02_coeffs_b} below) read
\begin{equation}
\left\{
\begin{array}{rl}
a_{0,2} = & - \left(\frac{2}{\gamma}\right)^{\frac{3}{2}} c \ s \ h_{1,0}^0\\
a_{1,1} = & \frac{2}{\gamma} s^{\frac{2}{\gamma}} \left[s^{1-\frac{2}{\gamma}} \delta_{\mathrm{BC}} h_{1,0}^0 + c \left((h_{1,0}^1)' + \delta_{\mathrm{BC}} \frac{s}{c} h_{1,0}^1\right) \right. \\
& \hspace{0.45cm} \left. - \frac{2}{\gamma} (I_{1-\frac{2}{\gamma}}-I_{-1-\frac{2}{\gamma}}) \left((h_{1,0}^0)' + \delta_{\mathrm{BC}} \frac{s}{c} h_{1,0}^0\right)\right]
\end{array}
\right.
\label{a02_a11}
\end{equation}
where all quantities are to be evaluated at $\theta = \theta_b$.

\subsection*{Step III: The growth rate $\bar{\omega}_0^2$}

Finally, we  put together the results from steps I and II, and obtain explicit expressions for the growth rate. Now, the resulting expression for $\bar{\omega}_0^2$ from \eqref{LinkBetweenEigenfrequencies} comes out as an expansion in $\bar{k}_y$, but  this is in fact physically most  relevant  since gravitational instability precisely occurs for small $\bar{k}_y$. More precisely, with the expansions \eqref{Def:hiaj_a0m_a1m} of $a_0$ and $a_1$, and expanding $\mathscr{C}$ with a similar notation as
\begin{equation}
\mathscr{C} =  \sum_{m=0}^\infty \mathscr{C}_m (\bar{k}_y)^m,
\label{Def:Cm}
\end{equation}
relation \eqref{LinkBetweenEigenfrequencies} becomes
\begin{equation}
\bar{\omega}_0^2 (\bar{k}_y) = \bar{\omega}_{0,1}^2 \ \bar{k}_y + \bar{\omega}_{0,2}^2 \ \bar{k}_y^2 + \mathcal{O}(\bar{k}_y^3)
\label{omega02_formula_2ndOrder}
\end{equation}
(it turns out that $\bar{\omega}_{0,0}^2=0$) where
\begin{subequations}
  \label{omega02_coeffs}
    \begin{empheq}[left={\empheqlbrace\,}]{align}
      & \bar{\omega}_{0,1}^2 = - \frac{a_{0,1}}{a_{1,0}} \label{omega02_coeffs_a} \\
      & \bar{\omega}_{0,2}^2 = \left(\frac{a_{0,2}}{a_{0,1}} - \frac{a_{1,1}}{a_{1,0}}\right) \omega_{0,1}^2 + \mathscr{C}_0 \ \omega_{0,1}^4. \label{omega02_coeffs_b}
    \end{empheq}
\end{subequations}
Then, with \eqref{a01_a10} on the one hand, and with \eqref{h010_and_h110} and \eqref{a02_a11} on the other hand, we respectively obtain the final result
\begin{equation}
\left\{
\begin{array}{rl}
\displaystyle
\bar{\omega}_{0,1}^2 = & - \sqrt{\tfrac{2}{\gamma} \left(1-\rho_b^\gamma\right)}\\
\displaystyle
\bar{\omega}_{0,2}^2 = & \displaystyle \left. \frac{4}{\gamma^2} \frac{c^2}{\frac{c}{s} y_Q'+ \delta_{\mathrm{BC}} y_Q}
\right\{ \frac{\gamma}{2} \left(1+\delta_{\mathrm{BC}}\right) s^{-\frac{2}{\gamma}} y_Q - \left(\frac{c}{s} y_P'+ \delta_{\mathrm{BC}} \ y_P \right) \ \Gamma\left(\tfrac{2}{\gamma}\right) \int_{\frac{\pi}{2}}^{\theta_b} \left(\frac{c}{s} y_Q'+y_Q\right) \frac{y_Q}{s} \dd\theta_1 \\
& \displaystyle
\left. + \left(\frac{c}{s} y_Q'+ \delta_{\mathrm{BC}} \ y_Q \right) \left[ \Gamma\left(\tfrac{2}{\gamma}\right) \int_{\frac{\pi}{2}}^{\theta_b} \! \left(\frac{c}{s} y_Q'+y_Q\right) \frac{y_P}{s} \dd\theta_1 + \frac{1}{c} \left(I_{-1-\frac{2}{\gamma}}-I_{1-\frac{2}{\gamma}}\right) + \frac{\gamma}{2} \mathscr{C}_0 \right] \right\}.
\end{array}
\right.
\label{omega02_coeffs_explicit}
\end{equation}
We remind the reader that in this expression all quantities outside integrals are evaluated at the boundary $\theta = \theta_b$, and we recall the following definitions: $\rho_b = \left(p_\mathrm{ext} / p_c\right)^{1/\gamma}$ where $p_c$ is the equilibrium pressure at the centre and $p_\mathrm{ext} \leq p_c$ is the external pressure\footnote{We exclude $p_\mathrm{ext} > p_c$ because in this case the condition \eqref{Def:xb} cannot be fulfilled and no proper equilibrium state exists.}; $\theta_b = \arcsin \rho_b^{\gamma/2}$; $s = \sin \theta$; $c = \cos \theta$; $y_P$ and $y_Q$ are defined in \eqref{Def:yP_yQ}; $\Gamma$ is the Gamma function; $\delta_{\mathrm{BC}} =0$ for Rigid BC and $\delta_{\mathrm{BC}} = 1$ for Free BC; $I_p$ is defined in \eqref{Def:Ip}; and $\mathscr{C}_0 = \mathscr{C}(\bar{k}_y = 0)$ is given by formula \eqref{CowlingTerm}.

Let us now give a couple of comments on this formula, beginning with the linear order (first line above).
First, this result is \textit{exact}, in the sense that we did not need the Cowling approximation to obtain it  as $\mathscr{C}$ does not show in the linear expansion. Second, it is interesting to note that this result is independent of the BC. Indeed, while $a_{0,1}$ and $a_{1,0}$ both depend on the BC, that dependence cancels out when we take their ratio.
Third, note that we have the following relations:
\begin{equation}
\int_0^{\bar{x}} \rho(s) \dd s = \frac{1}{1-\gamma} \int_0^{\bar{x}} (\rho^{\gamma-1})'' \dd s = \frac{1}{1-\gamma} (\rho^{\gamma-1})' = \sqrt{\tfrac{2}{\gamma} \left(1-\rho^\gamma\right)}
\end{equation}
where we used the Lane-Emden equation \eqref{LaneEmden_Dimensionless} in the first equality, in the second equality we used the fact that the profiles we are considering are flat at $\bar{x}=0$, and in the third equality we used expression \eqref{rho0prime} for the derivative of $\rho$. From this, we see that $\bar{\omega}_{0,1}^2$ in \eqref{omega02_coeffs_explicit} actually corresponds to the column density from the centre to the boundary of the slab, i.e. essentially to the mass (per unit area) of the structure.

\begin{figure}
\centering
\includegraphics[scale=0.6]{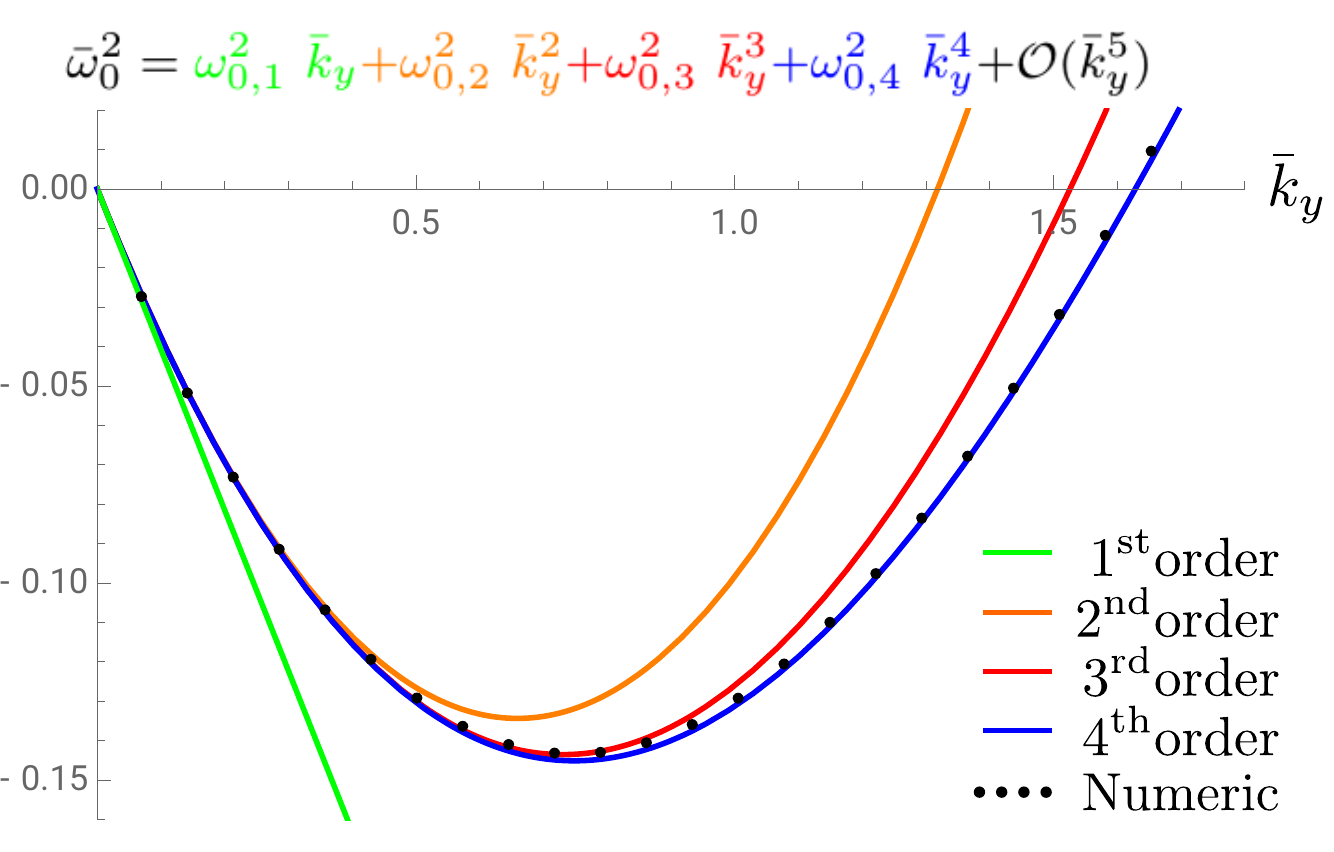}
\caption{At the top of this figure we reproduce the formula for the fundamental eigenfrequency $\bar{\omega}_0^2$ as an expansion in $\bar{k}_y$ up to the fourth order.
The color coding used in this expression corresponds to the different curves in the plot: green, orange, red and blue curves are respectively the linear, quadratic, cubic and quartic expansions. The numerical solution is shown in black dots. In this plot we naturally see that, as we increase the order, we increase the range of $\bar{k}_y$ over which the analytical result fits the numerical solution. In this example we chose Free BC, $\gamma=12$ and $\rho_b = 0.6$, and  the full unstable range (the range of $\bar{k}_y$ for which $\bar{\omega}_0^2$ is negative) is perfectly matched by the fourth order expansion. Note that the lower order expansions are all the better that $\rho_b$ is close to one (i.e. high external pressure), and the second order \eqref{omega02_formula_2ndOrder} becomes excellent for $\rho_b \gtrsim 0.8$ in general.}
\label{AccuracyOfTheVariousExpansions}
\end{figure}

About the quadratic order, we first point out that this result is \textit{exact} as long as we take $\mathscr{C}_0$ from its definition, namely the value of $\mathscr{C}$ defined in \eqref{Def:C} at $\bar{k}_y=0$, because we have computed exactly the term $a_1/a_0$ order by order in $\bar{k}_y$. It is only once we use expression \eqref{CowlingTerm} to express $\mathscr{C}_0$ explicitly that the result becomes approximate, since the latter was deduced using the Cowling approximation and the WKB method.
However, we find numerically that $\mathscr{C}_0$ is small compared to the other terms, all the more that $\rho_b$ approaches unity.
Second, we write $\bar{\omega}_{0,2}^2$ in \eqref{omega02_coeffs_explicit} in this manner in order to highlight the first term in the brackets, which dominates in the high pressure limit $\rho_b \rightarrow 1$.
Third, in figure~\ref{AccuracyOfTheVariousExpansions}, we show a typical example: the linear expansion is plotted in green, the quadratic expansion  in orange and the numerical solution  in black dots. As we can see, the quadratic formula is in excellent agreement with the numerical solution for small $\bar{k}_y$, namely for $\bar{k}_y$ from zero to a value close to $\bar{k}_{y,\mathrm{max}}$, the wavenumber of fastest growth (i.e.\ for $0 \leq \bar{k}_y \lesssim 0.5$ in the shown example). Therefore, from this quadratic formula one can already get good approximations of the maximum growth rate and the corresponding wavenumber, and fair approximations in the full unstable range, i.e. up to the critical wavenumber $\bar{k}_\mathrm{crit}$ at which the fundamental frequency $\bar \omega_0^2$ reaches zero again.

\subsection*{Beyond the quadratic formula}
\label{section:BeyondQuadraticFormula}

\begin{figure}
\centering
  \includegraphics[width=1.0\linewidth]{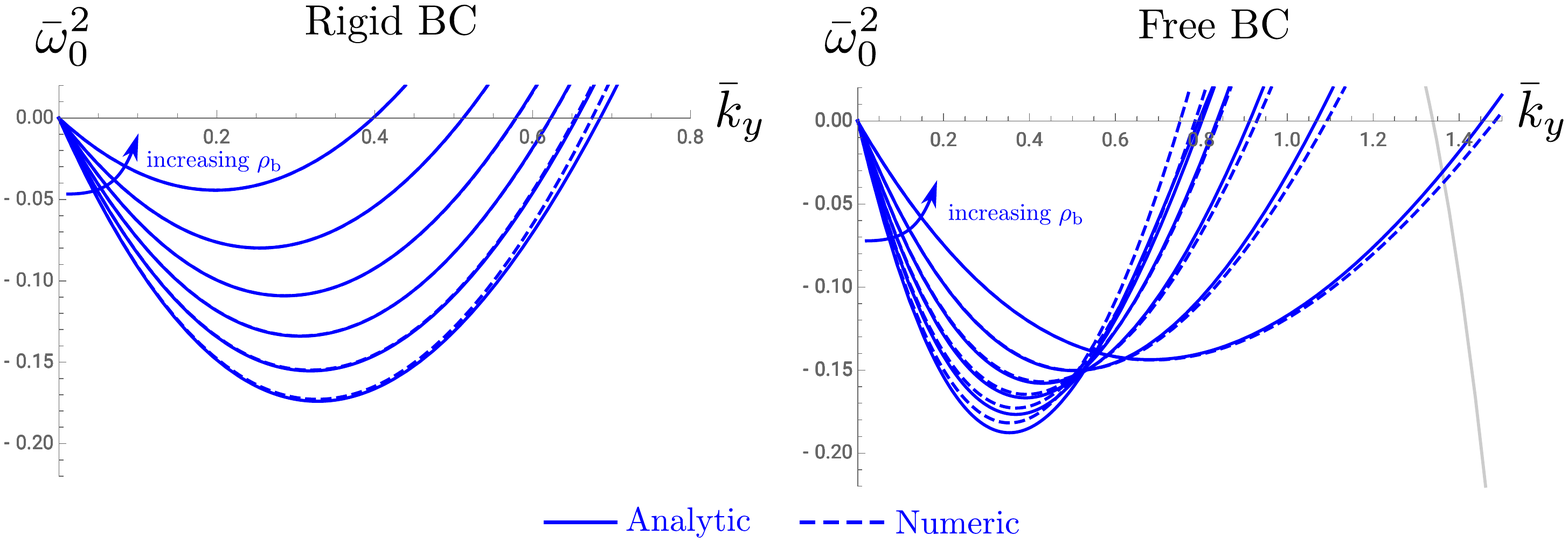}
\caption{Results fixing $\gamma = 1$ and varying $\rho_b$. Solid lines are the analytic solutions at fourth order, and dashed lines are the numerical solutions. In the left panel we use Rigid BC, while in the right panel we use Free BC. We indicate by an arrow how the slope at the origin varies as we increase the parameter $\rho_b$. Here we increase the values of $\rho_b$ from $0.4$ to $0.9$ by steps of $0.1$. Formally this evolution can be clearly seen with the expression of $\bar{\omega}_{0,1}^2$ in \eqref{omega02_coeffs_explicit}. The light grey branch in the right panel is explained in section \ref{section:BeyondQuadraticFormula}.}
\label{fig:ExampleOfResultsVaryRhob}
\end{figure}

\begin{figure}
\centering
  \includegraphics[width=1.0\linewidth]{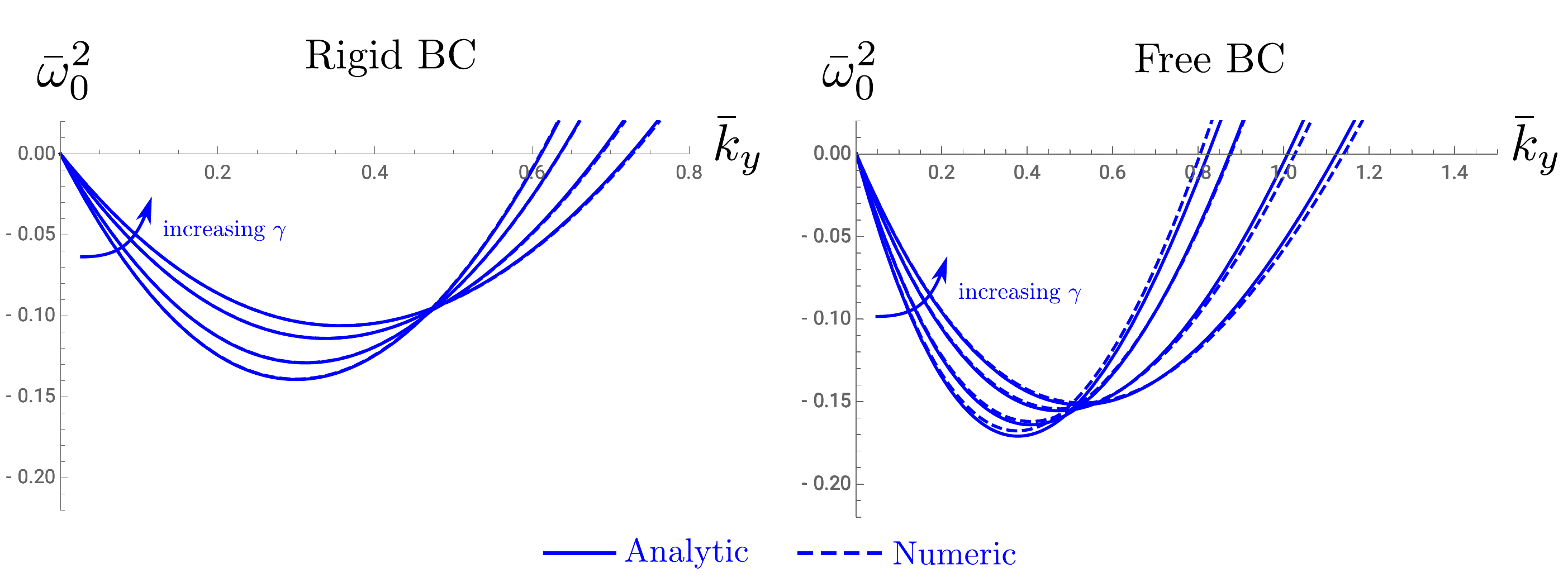}
\caption{Results fixing $\rho_b = 0.6$ and varying $\gamma$. Solid lines are the analytic solutions at fourth order, and dashed lines are the numerical solutions. In the left panel we use Rigid BC, while in the right panel we use Free BC. We indicate by an arrow how the slope at the origin varies as we increase the parameter $\gamma$. Here we plot the curves for  $\gamma = 0.5,1.5,3.5$ and $5$. Formally this evolution can be clearly seen with the expression of $\bar{\omega}_{0,1}^2$ in \eqref{omega02_coeffs_explicit}.}
\label{fig:ExampleOfResultsVaryGamma}
\end{figure}

An important aspect of the method presented in this paper is that it is systematic, in the sense that if need be, it is straightforward to increase the accuracy of expression \eqref{omega02_formula_2ndOrder} by increasing the order of the expansions\footnote{To be more precise, with formula \eqref{LinkBetweenEigenfrequencies} we have an `intrinsic' limitation, given by the accuracy of the Cowling approximation. To increase even further the accuracy, one may consider the  first three root-coefficient relations, rather than just the  first two as we did here. Doing so, we are limited by the accuracy of the Cowling approximation for one harmonic higher, and since that approximation is all the better than the harmonics are high, the total accuracy will indeed be improved. However, this improvement is at the cost of more cumbersome expressions, and given how precise the results are already here (cf. Figs.~\ref{fig:ExampleOfResultsVaryRhob} and \ref{fig:ExampleOfResultsVaryGamma}) it seems to us that this is not necessary in practice here. It could however turn out to be useful in more complicated systems (with magnetic field, non-static equilibrium, non-self-gravitating, etc.).}.
As a general rule, we find that in order to get excellent match between numerical and analytic results, we need to increase the order of the expansion in $\bar{k}_y$ as we lower the value of $\rho_b$. Indeed, for large values of $\rho_b$, typically larger than $0.8$, the second order can already be excellent, i.e. as good as the fourth order in Fig~\ref{AccuracyOfTheVariousExpansions}. In Figs~\ref{fig:ExampleOfResultsVaryRhob} and \ref{fig:ExampleOfResultsVaryGamma} we show that we obtain excellent agreement for a very wide range of parameters (varying $\gamma$ and $\rho_b$), but for this to happen on the full unstable range, we need to expand up to the fourth order. Now, the third order is already much more cumbersome than the second order, while it improves the accuracy by only typically several percents. Therefore quite a lot of effort is required to improve the accuracy of the quadratic formula, and whether or not the gain is worth the effort depends on the specific problem at hand. As exemplified by Fig~\ref{AccuracyOfTheVariousExpansions}, the quadratic formula already gives good results on the full unstable range, so in many cases it may constitute the best compromise.

The relevant scales for gravitational instability are $0 \leq \bar{k}_y \leq \bar{k}_{\mathrm{crit}}$, by definition of the critical wavenumber $\bar{k}_{\mathrm{crit}}$. As shown in figures~\ref{AccuracyOfTheVariousExpansions},\ref{fig:ExampleOfResultsVaryRhob} and \ref{fig:ExampleOfResultsVaryGamma}, over that range, our formulas are very efficient. However, let us note that they are generally not valid much beyond the unstable range. This is no surprise since, in the quartic formula  for instance, the function $\bar{\omega}_0^2$ is a fourth order polynomial in $\bar{k}_y$, and it has no lower bound for very large $\bar{k}_y$, which is obviously unphysical. This can be seen in the right panel of figure~\ref{fig:ExampleOfResultsVaryRhob}: the curve in light grey is part of the $\rho_b = 0.4$ solution, but we coloured that branch in light grey because it is outside the relevant domain. In the other curves of figures~\ref{AccuracyOfTheVariousExpansions},\ref{fig:ExampleOfResultsVaryRhob} and \ref{fig:ExampleOfResultsVaryGamma} this is not visible simply because it happens beyond the plotted regions.

\section{Conclusion}
\label{section:Conclusion}

We have presented a method to obtain analytic expressions for the growth rate of the gravitational instability of a planar pressure-confined static self-gravitating polytropic fluid. The results are in excellent agreement with the numerical resolution. Formula \eqref{omega02_coeffs_explicit} for the first and second order constitute the main explicit result of this paper. The strength of that formula lies in the fact that it allows us to understand explicitly how physical properties such as the column density, the pressure, the polytropic index, the boundary conditions, etc., control the behaviour of gravitational instability in polytropic self-gravitating sheets. In addition, as the approach we introduced here is perturbative, one may easily increase the accuracy of that formula by pushing the expansions further. The higher accuracy of the third and fourth order approximations is exhibited in Figs. \ref{AccuracyOfTheVariousExpansions}, \ref{fig:ExampleOfResultsVaryRhob} and \ref{fig:ExampleOfResultsVaryGamma}. Finally and most importantly, we emphasise that this method can naturally be extended to more complex systems (adding magnetic field, flow, expansion, other components, etc.). Indeed, all the steps we presented here are very general: the link \eqref{LinkBetweenEigenfrequencies} between the eigenvalues is universal, the Cowling approximation will always be excellent for high-order modes, the stable part of the spectrum is always constituted of large eigenvalues, and we solved the equations using only standard perturbative methods. Generalising the results of this paper is left for future work.

J.B.D. acknowledges financial support by the P2IO LabEx (ANR-10-LABX-0038) in the framework `Investissements d'Avenir' (ANR-11-IDEX-0003-01) managed by the French National Research Agency (ANR) when this work has been initiated. This work has been in part supported by MEXT Grant-in-Aid for Scientific Research on Innovative Areas No. 15K2173 (J.B.D.).

\appendix

\section{Details of step I (computing $\mathscr{C}$)}
\label{section:StepI}

In this Appendix we focus on the harmonics $\bar{\omega}_{n \geq 1}^2$, i.e.\ the stable part of the spectrum, by studying the dynamics in the Cowling approximation in order to derive an explicit expression for $\mathscr{C}$, namely \eqref{CowlingTerm}.

\subsection{Governing equations in the Cowling approximation}

In the Cowling approximation the eigenvalue problem is not \eqref{EigenvalueProblem} but 
\begin{equation}
\begin{array}{l}
\left\{
\begin{array}{l}
\hat{\rho}_1 = - (\rho_0 \hat{\xi}_x)' - \rho_0 i k_y \hat{\xi}_y\\
- \rho_0 \omega^2 \hat{\xi}_x = - c_a^2 \hat{\rho}_1' + \left(\hat{g}_0 - (c_a^2)'\right) \hat{\rho}_1\\
- \rho_0 \omega^2 \hat{\xi}_y = - i k_y c_a^2 \hat{\rho}_1
\end{array}
\right.
\end{array}
\label{EigenvalueProblem_Cowling}
\end{equation}
(where here $'=\frac{\dd}{\dd x}$) because we do not consider $\vec{g}_1$ which makes the linearised Poisson equation irrelevant. Then, using the  dimensionless parameters \eqref{DimensionlessVariables} defined above, it becomes
\begin{equation}
\frac{\dd}{\dd\bar{x}}
\left(
\begin{array}{c}
\psi\\
\mathcal{R}
\end{array}
\right)
=
\left(
\begin{array}{cccc}
0 & \frac{\bar{k}_y^2}{\bar{\omega}^2} \rho(\bar{x}) - \rho(\bar{x})^{2-\gamma}\\
\frac{\bar{\omega}^2}{\rho(\bar{x})} & 0
\end{array}
\right)
\left(
\begin{array}{c}
\psi\\
\mathcal{R}
\end{array}
\right).
\label{EqnInxbar_Cowling}
\end{equation}
Changing to the $\rho$ variable yields
\begin{equation}
\frac{\dd}{\dd\rho}
\left(
\begin{array}{c}
\psi\\
\mathcal{R}
\end{array}
\right)
=
- \sqrt{\frac{\gamma}{2}} \frac{\rho^{\gamma-2}}{\sqrt{1-\rho^\gamma}}
\left(
\begin{array}{cccc}
0 & \frac{\bar{k}_y^2}{\bar{\omega}^2} \rho - \rho^{2-\gamma}\\
\frac{\bar{\omega}^2}{\rho} & 0
\end{array}
\right)
\left(
\begin{array}{c}
\psi\\
\mathcal{R}
\end{array}
\right),
\label{EqnInrho_Cowling}
\end{equation}
and  with the $\theta$ variable we finally get
\begin{equation}
\frac{\dd}{\dd\theta}
\left(
\begin{array}{c}
\psi\\
\mathcal{R}
\end{array}
\right)
=
- \sqrt{\frac{2}{\gamma}}
\left(
\begin{array}{cccc}
0 & \frac{\bar{k}_y^2}{\bar{\omega}^2} s - s^{2/\gamma-1}\\
\bar{\omega}^2 s^{1-4/\gamma} & 0
\end{array}
\right)
\left(
\begin{array}{c}
\psi\\
\mathcal{R}
\end{array}
\right).
\label{EqnIntheta_Cowling}
\end{equation}
This equation corresponds to \eqref{EqnIntheta} but in the Cowling approximation. Now, differentiating the second line of \eqref{EqnIntheta_Cowling} and plugging it in the first, it is straightforward to get the following scalar equation on $\mathcal{R}$ 
\begin{equation}
\mathcal{R}'' + \left(\frac{4}{\gamma}-1\right) \frac{c}{s} \ \mathcal{R}' + \frac{2}{\gamma} \left(s^{-1-\frac{2}{\gamma}} \bar{\omega}^2 - s^{1-\frac{4}{\gamma}} \bar{k}_y^2\right) \mathcal{R} = 0,
\label{EqnOnP_Cowling}
\end{equation}
where, and from now on, $'=\frac{\dd}{\dd\theta}$. We prefer to work with the equation on $\mathcal{R}$ because it turns out to be simpler than that satisfied by $\psi$. We now need just to solve \eqref{EqnOnP_Cowling}, and from the obtained expression of $\mathcal{R}$, we will be able get $\psi$ by inverting the second line of \eqref{EqnIntheta_Cowling}, i.e. 
\begin{equation}
\psi = - \frac{1}{\bar{\omega}^2} \sqrt{\frac{\gamma}{2}} s^{\frac{4}{\gamma}-1} \frac{\dd \mathcal{R}}{\dd\theta}.
\label{PsiAsFunctionOfPprime_Cowling}
\end{equation}

\subsection{The eigenfunctions $\psi$ and $\mathcal{R}$ for $n \geq 1$}

In order to solve \eqref{EqnOnP_Cowling}, it is convenient to change  variables in the following way:
\begin{equation}
\mathcal{R} = (\sin \theta)^{\beta} \ u
\label{ChangeVariablePtoU}
\end{equation}
where  the chosen value of the constant $\beta$ is such  that in the equation satisfied by $u$ the term with a first derivative vanishes, i.e.
\begin{equation}
\beta = \frac{1}{2} - \frac{2}{\gamma}.
\end{equation}
Then we have
\begin{equation}
u'' +(\bar{\omega}^2 q - f) u = 0
\label{SchrodingerEquation}
\end{equation}
where
\begin{equation}
\left\{
\begin{array}{l}
q(\theta) = \frac{2}{\gamma} s^{-\frac{2}{\gamma}}\\
f(\theta) = \frac{2}{\gamma} \bar{k}_y^2 s^{2-\frac{4}{\gamma}} + \left(\frac{2}{\gamma} - \frac{1}{2}\right)\left(\frac{2}{\gamma} - \frac{3}{2}\right) s^{-2} - \left(\frac{2}{\gamma} - \frac{1}{2}\right)^2.
\end{array}
\right.
\label{Def_q_and_f}
\end{equation}
Now, as we are looking for the high order modes (i.e. not the fundamental), we know that the values of interest of the parameter $\bar{\omega}$, namely the eigenvalues, are large. Therefore, let us treat the quantity $1/\bar{\omega}$ as a small parameter, in which case we
may write explicitly the solution of an equation of the classical form \eqref{SchrodingerEquation} using the WKB method. Following the presentation of \cite{Holmes95}, we set
\begin{equation}
\epsilon \equiv \frac{1}{\bar{\omega}}
\end{equation}
and look for solutions in the form
\begin{equation}
u(\theta) \sim \left[u_0(\theta) + \epsilon u_1(\theta) + \epsilon^{2} u_2(\theta) + \dots\right] \exp \left(\frac{A(\theta)}{\epsilon}\right).
\end{equation}
Plugging this  into \eqref{SchrodingerEquation} and identifying by powers of $\epsilon$ gives a hierarchy of equations that can be solved one after the other. Note that if we take only the first  WKB order, the spectrum $\left\{ \bar{\omega}^2_n \right\}$ at this level of approximation does not depend on $\bar{k}_y$. Therefore we have to develop $u(\theta)$ up to the second order, which is why we go up to the $\epsilon^{2} u_2$ term in the above expression. We find then that the general solution of \eqref{SchrodingerEquation} is
\begin{equation}
u(\theta) \sim q^{-\frac{1}{4}} \left[a \left(1 + \frac{i K}{\bar{\omega}}\right) \exp \left(i \bar{\omega} \int_{\frac{\pi}{2}}^\theta \sqrt{q} \dd x \right) + b \left(1 - \frac{i K}{\bar{\omega}}\right) \exp \left(-i \bar{\omega} \int_{\frac{\pi}{2}}^\theta \sqrt{q} \dd x \right) \right]
\label{WKB_intermediaire}
\end{equation}
where $a$ and $b$ are constants, possibly complex, determined by the boundary (or symmetry) conditions, and
\begin{equation}
K(\theta) \equiv \frac{1}{2} \int_{\frac{\pi}{2}}^\theta \frac{1}{\sqrt{q}} \left(\frac{5}{16} \left(\frac{q'}{q}\right)^2 - \frac{1}{4} \frac{q''}{q} - f \right) \dd x.
\label{K_intermediaire}
\end{equation}
Using  definitions \eqref{Def_q_and_f} of $q$ and $f$ and definition \eqref{Def:Ip} of $I_p$, expression \eqref{K_intermediaire} becomes
\begin{equation}
K(\theta) = \frac{\gamma^{-\frac{3}{2}}}{8 \sqrt{2}} \left\{(3-\gamma)(5-\gamma) \ I_{\tfrac{1}{\gamma}}(\theta) - (3-\gamma)(5-3 \gamma) \ I_{\tfrac{1}{\gamma}-2}(\theta) - 8 \gamma \ I_{2-\tfrac{3}{\gamma}}(\theta) \ \bar{k}_y^2\right\}.
\label{K_Explicit}
\end{equation}
Then, with relation \eqref{PsiAsFunctionOfPprime_Cowling}, the symmetry condition $\psi(\pi/2) = 0$ becomes $\mathcal{R}'(\pi/2) = 0$ and given the change of variables \eqref{ChangeVariablePtoU}, we have $\mathcal{R}' = \sin(\theta)^\beta (u'+\beta \cot (\theta) u)$, so that this symmetry condition translates into
\begin{equation}
u'\left(\tfrac{\pi}{2}\right) = 0.
\end{equation}
Therefore $a=b$ in \eqref{WKB_intermediaire}. Finally, rewriting the complex number $1 \pm i K/\bar{\omega}$ in exponential form ($K$ is real), we can rewrite the solution \eqref{WKB_intermediaire} as
\begin{equation}
u(\theta) = a \sqrt{1 + \frac{K^2}{\bar{\omega}^2}} q^{-\frac{1}{4}} \cos \left(\bar{\omega} \int_{\frac{\pi}{2}}^\theta \sqrt{q} \dd\theta_1 + \varphi \right)
\label{WKB_solution}
\end{equation}
where the phase is given by
\begin{equation}
\varphi = \arctan \left(\frac{K}{\bar{\omega}}\right).
\end{equation}
To help get an intuition of this expression, note that, going back the various changes of variables above ($\theta \leftrightarrow \rho$ and $\rho \leftrightarrow \bar{x}$), we have
\begin{equation}
\int_{\frac{\pi}{2}}^\theta \sqrt{q} \dd\theta_1 = - \int_0^{\bar{x}} \sqrt{\rho(\bar{x})^{1-\gamma}} \dd\bar{x},
\end{equation}
so that for example in the isothermal case $\gamma = 1$ this factor is in fact simply equal to (minus) the position $\bar{x}$.

Finally, plugging relation \eqref{WKB_solution} into \eqref{ChangeVariablePtoU} directly gives the expression for $\mathcal{R}$, and with relation \eqref{PsiAsFunctionOfPprime_Cowling} the function $\psi$ is obtained. In figure~\ref{Cowling_Eigenfunctions_And_Eigenvalues} we compare them with the numerical resolution of \eqref{EqnIntheta_Cowling}.

\subsection{The eigenvalues $\bar{\omega}_n^2$ for $n \geq 1$}

The expressions for $\mathcal{R}$ and $\psi$ that we just obtained are parameterised by $\bar{\omega}^2$, since they are essentially given by \eqref{WKB_solution}. As we plug them into the boundary condition \eqref{BC2} we get\footnote{We can indeed divide by $\cos \left(\bar{\omega} \int_{\frac{\pi}{2}}^\theta \sqrt{q} \dd\theta_1 + \varphi \right)$ because it must be non zero, otherwise the BC \eqref{BC2} is not satisfied.}
\begin{equation}
\tan \left(\bar{\omega} \int_{\frac{\pi}{2}}^{\theta_b} \sqrt{q} \dd\theta_1 + \varphi \right) = \Phi
\label{TanEqualsPhiN}
\end{equation}
where
\begin{equation}
\Phi \equiv \frac{1}{\bar{\omega}} \frac{K K' + \left(\frac{1}{2}-\frac{3}{2 \gamma}\right) \frac{c}{s} - \sqrt{\frac{2}{\gamma}} \Delta_\mathrm{BC} s^{1-\frac{4}{\gamma}} \bar{\omega}^2 \left(\bar{\omega}^2 + K^2\right)}{K' + \sqrt{\frac{2}{\gamma}} s^{-\frac{1}{\gamma}} \left(\bar{\omega}^2 + K^2\right)}.
\label{GrandPhiN}
\end{equation}
This is a rather complicated equation on $\bar{\omega}$, and there is little hope to find its solutions in closed form.
Therefore we solve it perturbatively by looking for solutions in the form
\begin{equation}
\bar{\omega}_n = \bar{\omega}_{(1)} n + \bar{\omega}_{(0)} + \bar{\omega}_{(-1)} \frac{1}{n} + \dots
\label{Solving_Harmonics_Perturbatively}
\end{equation}  
Since these correspond to the high order modes, $n$ is a large parameter, which is not convenient to perform Taylor expansions. Therefore we rewrite \eqref{TanEqualsPhiN} in a way such that we may simply use Taylor expansions near $0$, i.e. developments in $1/n$ only. For this, we first apply $\arctan$, keeping in mind that this brings an $n \pi$ term. Then the choice of BC matters. For rigid BC ($\Delta_\mathrm{BC} = 0$) we have $\Phi \sim 1/n^3$ for large $n$, so we can Taylor expand, but for Free BC ($\Delta_\mathrm{BC} \neq 0$) we have $\Phi \sim n$, so that we first use the identity\footnote{This is what brings a difference in the two types of BC in the solution \eqref{Spectrum_Cowling} ($\bar{\omega}_n^2$ contains $n^2$ for Rigid BC while it is $(n-1/2)^2$ for Free BC).} $\arctan(x) = \frac{\pi}{2} \times \mathrm{sgn}(x) - \arctan(\frac{1}{x})$, where $\mathrm{sgn}(x)$ is the sign of $x$, to work only with $1/n$ indeed. Finally, we also divide the whole equation by $n$, put everything on the left hand side and define
\begin{equation}
G \equiv \frac{\bar{\omega}_n^2}{n^2}
\end{equation}
in order to factor out the $n$ dependence. Doing so, we rewrite \eqref{TanEqualsPhiN} for Rigid BC as
\begin{equation}
\mathrm{sgn} (\Phi) \pi - \sqrt{\tfrac{2 G}{\gamma}} I_{-\frac{1}{\gamma}}(\theta_b) - \frac{1}{n} \arctan \left(\frac{1}{n} \frac{K}{\sqrt{G}}\right) + \frac{1}{n} \arctan \left[\frac{1}{n^3 \sqrt{G}} \frac{ \frac{c}{s} \left(\frac{1}{2}-\frac{3}{2 \gamma}\right)  \left(G + \frac{K^2}{n^2}\right) + \frac{K K'}{n^2}}{\sqrt{\frac{2}{\gamma}} s^{-\frac{1}{\gamma}} \left(G + \frac{K^2}{n^2}\right) + \frac{K'}{n^2}} \right] = 0,
\end{equation}
and for Free BC as
\begin{equation}
\begin{array}{l}
\displaystyle
\mathrm{sgn} (\Phi) \pi \left(1 - \tfrac{1}{2 n}\right) - \sqrt{\tfrac{2 G}{\gamma}} I_{-\frac{1}{\gamma}}(\theta_b) - \frac{1}{n} \arctan \left(\frac{1}{n} \frac{K}{\sqrt{G}}\right) - \frac{1}{n} \arctan \left[\frac{\sqrt{G}}{n} \left(\sqrt{\tfrac{2}{\gamma}} s^{-\frac{1}{\gamma}} \left(G + \frac{K^2}{n^2}\right) + \frac{K'}{n^2}\right) \right.\\
\displaystyle \left. \hspace{2.25cm} \times \left(- \sqrt{\tfrac{2}{\gamma}} \Delta_\mathrm{BC} s^{1-\frac{4}{\gamma}} G \left(G + \frac{K^2}{n^2}\right) + \frac{1-\tfrac{3}{\gamma}}{2 n^2} \frac{c}{s} \left(G + \frac{K^2}{n^2}\right) + \frac{K K'}{n^4}\right)^{-1} \right] = 0.
\end{array}
\end{equation}
We may now Taylor expand the functions on the left hand sides, and then identifying by orders of $1/n$ we get a system of equations for the coefficients in the expansion \eqref{Solving_Harmonics_Perturbatively}. Solving this system and rearranging terms, we finally get that the solution up to order $\mathcal{O}\left(n^{-2}\right)$ is
\begin{equation}
\bar{\omega}_n^2 = 
\frac{A \ \left(n-\tfrac{1}{2} \delta_\mathrm{BC}\right)^2 + B \ \bar{k}_y^2 + C + D_\mathrm{BC}}{I_{-\frac{1}{\gamma}}(\theta_b)}
\label{Spectrum_Cowling}
\end{equation}
where we used the notation $\delta_\mathrm{BC}$ defined in \eqref{delta_BC} to avoid separating the cases, and the coefficients $A,B,C$ and $D_\mathrm{BC}$ were defined in \eqref{Coeffs_A_B_C_D}.

\subsection{Expression for $\mathscr{C}$}

We are now finally ready to compute the series $\mathscr{C}$ given by \eqref{Def:C}. The pleasant surprise is that the $\bar{\omega}^2_n$'s given by \eqref{Spectrum_Cowling} have simple quadratic dependencies in $n$, so that we may obtain a closed form for the expression of $\mathscr{C}$, by using the identity
\begin{equation}
\sum_{n=1}^\infty \frac{1}{\left(n-\frac{1}{2}\right)^2 + b^2} = \frac{\pi \tanh(b \pi)}{2 b}
\end{equation}
for the Free BC ($\delta_\mathrm{BC} = 1$), and
\begin{equation}
\sum_{n=1}^\infty \frac{1}{n^2 + b^2} = - \frac{1}{2 b^2} + \frac{\pi \coth(b \pi)}{2 b}
\end{equation}
for the Rigid BC case ($\delta_\mathrm{BC} = 0$). In order to show a single expression combining the two types of BC, rather than having to separate the cases inelegantly, we again used the notation $\delta_\mathrm{BC}$ to write $\mathscr{C}$ in the form \eqref{CowlingTerm}.

\section{Details about step II (computing $a_1/a_0$)}
\label{section:StepII}

In this Appendix we show how to compute $a_0$ and $a_1$, and in particular how expressions \eqref{h010_and_h110}, \eqref{a01_a10} and \eqref{a02_a11} are obtained.

\subsection{Equations for the $h_a^0$'s and $h_a^1$'s}

From the relation \eqref{ak} defining the $a_k$'s, it turns out that we only need the $h_a^0$'s and the $h_a^1$'s, not higher order terms in the expansion in $\bar{\omega}^2$ of $\mathsfbi{H}$ in order to compute $a_0$ and $a_1$.
To find them, let us derive the differential equations they satisfy. First we obtain the equation satisfied by the matrix $\mathsfbi{H}$, by plugging \eqref{PGy_AvecH} and \eqref{PsiGx_WithInverseB} in the equation formed by the first two rows of \eqref{EqnIntheta}. Then, treating $\mathcal{R}(\pi/2)$ and  $\mathcal{G}_y(\pi/2)$ as independent variables, we send each coefficient of that matrix equation to zero, which gives four equations. Finally, plugging definition \eqref{Def:hia} of the $h_a^i$'s, we can identify by powers of $\bar{\omega}^2$ to get the following equations.
The zeroth power gives the equations for $h_1^0$ and $h_3^0$, namely
\begin{subequations}
  \label{syst_0_for_1and3}
    \begin{empheq}[left={\empheqlbrace\,}]{align}
      & (h_1^0)'' + \frac{c}{s} \left(\tfrac{4}{\gamma}-1\right) (h_1^0)' + \frac{2}{\gamma} \left(1-\bar{k}_y^2 s^{2-\frac{4}{\gamma}}\right) h_1^0 + \frac{2}{\gamma} \frac{c}{s} \frac{1}{\bar{k}_y^2} (h_3^0)' = 0 \label{syst_0_for_1and3_a} \\
      & (h_3^0)'' + \frac{c}{s} \left(\tfrac{2}{\gamma}-1\right) (h_3^0)' - \frac{2}{\gamma} \bar{k}_y^2 s^{2-\frac{4}{\gamma}} h_3^0 - \frac{2}{\gamma} \bar{k}_y^2 h_1^0 = 0
	  &  \label{syst_0_for_1and3_b}
    \end{empheq}
\end{subequations}
and, with the substitution $(1 \rightarrow 2)$ and $(3 \rightarrow 4)$ in the subscripts, one obtains the equations for $h_2^0$ and $h_4^0$.
Then the powers $i \geq 1$ give for $h_1^i$ and $h_3^i$,
\begin{subequations}
  \label{syst_i_for_1and3}
    \begin{empheq}[left={\empheqlbrace\,}]{align}
      & (h_1^i)'' + \frac{c}{s} \left(\tfrac{4}{\gamma}-1\right) (h_1^i)' + \frac{2}{\gamma} \left(1-\bar{k}_y^2 s^{2-\frac{4}{\gamma}}\right) h_1^i + \frac{2}{\gamma} \frac{c}{s} \frac{1}{\bar{k}_y^2} (h_3^i)' + \frac{2}{\gamma} s^{-\frac{2}{\gamma}} h_1^{i-1} = 0 \label{syst_i_for_1and3_a} \\
      & (h_3^i)'' + \frac{c}{s} \left(\tfrac{2}{\gamma}-1\right) (h_3^i)' - \frac{2}{\gamma} \bar{k}_y^2 s^{2-\frac{4}{\gamma}} h_3^i - \frac{2}{\gamma} \bar{k}_y^2 h_1^i = 0
	  &  \label{syst_i_for_1and3_b}
    \end{empheq}
\end{subequations}
and, with the substitution $(1 \rightarrow 2)$ and $(3 \rightarrow 4)$ in the subscripts, one obtains the equations for $h_2^i$ and $h_4^i$.
Note that, as stated above, in order to get $\bar{\omega}_0^2$ with \eqref{LinkBetweenEigenfrequencies}, we only need $h_a^0$ and $h_a^1$, so that in fact only the case $i=1$ above is used in the following. However, if one is also interested in the full expression of the eigenfunctions, then in principle all the $h_a^i$'s are required, and it turns out that solving for $i=1$ or generalising to any $i \geq 1$ is immediate, so we might as well treat the more general case directly. 

\subsection{Solving for $h_1^0$ and $h_3^0$}
\label{section:SolvingForh01Andh03}

Let us focus first on the system \eqref{syst_0_for_1and3}, i.e. on the pair $(h_1^0,h_3^0)$. A most natural way of pursuing further is to express $h_1^0$ in terms of $h_3^0$ and its derivatives from the equation \eqref{syst_0_for_1and3_b} and plug that into \eqref{syst_0_for_1and3_a}. This gives an equation for $h_3^0$ that we could try and solve. Had we managed to solve it, we would be able to deduce $h_1^0$ using \eqref{syst_0_for_1and3_b}.
This approach however requires to deal with complicated fourth order differential equations.

Instead, let us remember that  pressure balances gravity on small scales, giving rise to acoustic oscillations. We thus expect gravitational instability to be triggered by large scales perturbations, i.e. for small $\bar{k}_y$. It is therefore natural to look for the $h_a^i$'s by expanding them in terms of $\bar{k}_y$ around $0$. As we will see right now, this actually enables us to reduce the problem to just second order equations. Most remarkably, these equations possess well known closed form solutions.
Indeed, let us now expand the $h_a^i$'s as
\begin{equation}
h_a^i = \sum_{j=0}^\infty h_{a,j}^i \left(\bar{k}_y^2\right)^j \hspace{1cm} \mathrm{where} \ h_{a,j}\equiv h_{a,j}^i(\theta).
\label{h_a_i_InTermsOf_ky2}
\end{equation}
Then, identifying in powers of $\bar{k}_y^2$ in \eqref{syst_0_for_1and3_a} gives
\begin{subequations}
  \label{system1}
    \begin{empheq}[left={\empheqlbrace\,}]{align}
      & (h_{3,0}^0)' = 0,
        \label{system1a} \\
      & (h_{1,0}^0)'' + \frac{c}{s} \left(\tfrac{4}{\gamma}-1\right) (h_{1,0}^0)' + \frac{2}{\gamma} h_{1,0}^0 + \frac{2}{\gamma} \frac{c}{s} (h_{3,1}^0)' = 0,
        \label{system1b}\\
      & \mathrm{and \ for} \ j \geq 1 : \nonumber \\
      & (h_{1,j}^0)'' + \frac{c}{s} \left(\tfrac{4}{\gamma}-1\right) (h_{1,j}^0)' + \frac{2}{\gamma} h_{1,j}^0 - \frac{2}{\gamma}  s^{2-\frac{4}{\gamma}} h_{1,j-1}^0 + \frac{2}{\gamma} \frac{c}{s} (h_{3,j+1}^0)' = 0, \label{system1c}
    \end{empheq}
\end{subequations}
and in \eqref{syst_0_for_1and3_b} it gives
\begin{subequations}
  \label{system2}
    \begin{empheq}[left={\empheqlbrace\,}]{align}
      & (h_{3,0}^0)'' + \frac{c}{s} \left(\tfrac{2}{\gamma}-1\right) (h_{3,0}^0)' = 0,
        \label{system2a} \\
      & \mathrm{and \ for} \ j \geq 1 : \nonumber \\
	  & (h_{3,j}^0)'' + \frac{c}{s} \left(\tfrac{2}{\gamma}-1\right) (h_{3,j}^0)' - \frac{2}{\gamma} h_{1,j-1}^0 - \frac{2}{\gamma}  s^{2-\frac{4}{\gamma}} h_{3,j-1}^0 = 0. \label{system2b}
    \end{empheq}
\end{subequations}
In the following, we will need constraints to determine the constants of integration when solving these equations. We get the constants using the fact that from the definition \eqref{PGy_AvecH} of $\mathsfbi{H}$, 
\begin{equation}
\mathsfbi{H}\left(\frac{\pi}{2}\right) = 1 \! \! 1
\label{H_piSur2}
\end{equation}
and given the relation \eqref{PsiGx_WithInverseB} between $\mathsfbi{H}$ and $(\psi,\mathcal{G}_x)$ together with the symmetry conditions \eqref{SymmetryConditions_newVariables}, we take
\begin{equation}
\mathsfbi{H}'\left(\frac{\pi}{2}\right) = 0.
\label{Hprime_piSur2}
\end{equation}
Let us now analyse the equations from \eqref{system1} and \eqref{system2} as follows.

First, consider \eqref{system1a} and \eqref{system2a}:  equation \eqref{system2a} does not add any new information here, since it is automatically satisfied due to \eqref{system1a}. At least it is a consistency check. Then \eqref{system1a} with \eqref{H_piSur2} directly gives
\begin{equation}
h_{3,0}^0(\theta) = 0
\end{equation}
for all $\theta$.

Second, consider \eqref{system1b} and \eqref{system2b} with $j=1$: summing these  together gives
\begin{equation}
(h_{1,0}^0 + h_{3,1}^0)'' + \frac{c}{s} \left(\frac{4}{\gamma}-1\right) (h_{1,0}^0 + h_{3,1}^0)' = 0.
\label{eqIntermediaire1}
\end{equation}
This is simply a linear homogeneous first order differential equation for the function $(h_{1,0}^0 + h_{3,1}^0)'$. Its solution reads
\begin{equation}
(h_{1,0}^0 + h_{3,1}^0)' = y_0 \ (\sin \theta)^{1-\frac{4}{\gamma}}
\label{SolutionHomogene1}
\end{equation}
where $y_0$ is a constant. Now with \eqref{Hprime_piSur2} we have that $(h_{1,0}^0)'(\frac{\pi}{2}) = 0$ and $(h_{3,1}^0)'(\frac{\pi}{2})=0$, so that in fact $y_0 = 0$, and \eqref{SolutionHomogene1} becomes trivial to integrate. The constant of integration is given by \eqref{H_piSur2}, and we finally get that for all $\theta$
\begin{equation}
h_{3,1}^0(\theta) = 1 - h_{1,0}^0(\theta).
\label{h031}
\end{equation}
We may now plug this expression of $h_{3,1}^0$ into \eqref{system1b} to get an equation on $h_{1,0}^0$ only. This yields 
\begin{equation}
\mathcal{L}[h_{1,0}^0] = 0
\label{EquDiff_h010}
\end{equation}
where the operator $\mathcal{L}$ is such that
\begin{equation}
\mathcal{L}[y] \equiv y''+ \left(\frac{2}{\gamma}-1\right) \frac{\cos \theta}{\sin \theta} y' + \frac{2}{\gamma} y.
\label{Def:L}
\end{equation}
The letter $\mathcal{L}$ stands for `Legendre' because when $\mathcal{L}[y] = 0$, the function $z = (\sin \theta)^{\frac{1}{\gamma}-1} y$ satisfies
\begin{equation}
z''+ \frac{\cos \theta}{\sin \theta} z' + \left\{\lambda \left(\lambda+1\right) - \frac{\mu^2}{\sin^2 \theta}\right\} z = 0
\end{equation}
where $\lambda = \frac{1}{\gamma}$ and $\mu=1-\frac{1}{\gamma}$, which is the associated Legendre equation, with solutions $P_\lambda^\mu$ and $Q_\lambda^\mu$. In other words, the general solution of \eqref{EquDiff_h010} is
\begin{equation}
h_{1,0}^0(\theta) = A_{1,0}^0 \ y_P(\theta) + B_{1,0}^0 \ y_Q(\theta)
\label{h010}
\end{equation}
where it is useful to define
\begin{equation}
\left\{
\begin{array}{l}
y_P(\theta) = (\sin \theta)^{1-\frac{1}{\gamma}} P_{\frac{1}{\gamma}}^{1-\frac{1}{\gamma}}(\cos \theta)\\
y_Q(\theta) = (\sin \theta)^{1-\frac{1}{\gamma}} Q_{\frac{1}{\gamma}}^{1-\frac{1}{\gamma}}(\cos \theta)\\
\end{array}
\right.
\label{Def:yP_yQ_InAppendix}
\end{equation}
because all the following results may  naturally be expressed in terms of these two functions. The coefficients $A_{1,0}^0$ and $B_{1,0}^0$ are constants, and using relations \eqref{H_piSur2} and \eqref{Hprime_piSur2}, we get that 
\begin{equation}
A_{1,0}^0 = 0 \hspace{0.5cm} \mathrm{and} \hspace{0.5cm} B_{1,0}^0 = [Q_{\frac{1}{\gamma}}^{1-\frac{1}{\gamma}}(0)]^{-1},
\label{A010_B010}
\end{equation}
hence the expression for $h_{1,0}^0$ in \eqref{h010_and_h110}. Now that   $h_{1,0}^0$ is obtained,  $h_{3,1}^0$ simply follows from \eqref{h031}.

Finally, consider \eqref{system1c} for a given $j \geq 1$ and \eqref{system2b} with $j+1$: summing these together gives
\begin{equation}
(h_{1,j}^0 + h_{3,j+1}^0)'' + \frac{c}{s} \left(\frac{4}{\gamma}-1\right) (h_{1,j}^0 + h_{3,j+1}^0)' = \frac{2}{\gamma} s^{2-\frac{4}{\gamma}} \left[h_{1,j-1}^0 + h_{3,j}^0\right].
\label{Recurrence_h03jplus1}
\end{equation}
At this point, it is important to notice the two things. First, the left hand side above is the same as  in \eqref{eqIntermediaire1}, so that the solution of the homogeneous equation is simply like \eqref{SolutionHomogene1}. Second, the right hand side is a known function, by iteration. Indeed, the solutions \eqref{h031} and \eqref{h010} we obtained above initialise the iterative process \eqref{Recurrence_h03jplus1}. Hence \eqref{Recurrence_h03jplus1} is simply a linear first order differential equation with a source term. Applying the method of variation of parameters, the general solution is
\begin{equation}
(h_{1,j}^0 + h_{3,j+1}^0)' = (\sin \theta)^{1-\frac{4}{\gamma}} \frac{2}{\gamma} \int_{\frac{\pi}{2}}^\theta \sin \theta_1 \left[h_{1,j-1}^0 + h_{3,j}^0\right] \dd\theta_1
\end{equation}
where the conditions $(h_{1,j}^0)'(\frac{\pi}{2})=0$ and $(h_{3,j+1}^0)'(\frac{\pi}{2})=0$ from \eqref{Hprime_piSur2} have been used. Integrating once more gives then
\begin{equation}
h_{3,j+1}^0(\theta) = - h_{1,j}^0(\theta) + \int_{\frac{\pi}{2}}^\theta (\sin \theta_1)^{1-\frac{4}{\gamma}} \frac{2}{\gamma} \int_{\frac{\pi}{2}}^{\theta_1} \sin \theta_2 \left[h_{1,j-1}^0 + h_{3,j}^0\right] \dd\theta_1 \dd\theta_2
\label{h03jplus1}
\end{equation}
where we used the conditions $(h_{1,j}^0)(\frac{\pi}{2})=0$ and $(h_{3,j+1}^0)(\frac{\pi}{2})=0$ from \eqref{H_piSur2}.

We may now insert this expression of $h_{3,j+1}^0$ back into \eqref{system1c} and get an equation on $h_{1,j}^0$ only. It is the inhomogeneous equation
\begin{equation}
\mathcal{L}[h_{1,j}^0] = \mathcal{S}_{1,j}^0
\label{eqIntermediaire2}
\end{equation}
where the source term on the right hand side is 
\begin{equation}
\mathcal{S}_{1,j}^0 = \frac{2}{\gamma}  s^{2-\frac{4}{\gamma}} h_{1,j-1}^0 - \frac{4}{\gamma^2} \ c \ s^{-\frac{4}{\gamma}} \int_{\frac{\pi}{2}}^\theta \sin \theta_1 \left[h_{1,j-1}^0 + h_{3,j}^0\right] \dd\theta_1.
\end{equation}
We obtain  the explicit solution of \eqref{eqIntermediaire2} below, as we generalise it.

\subsection{Generalisation: Solving for all the $h_{a,j}^i$'s}

Because the systems of equations \eqref{syst_0_for_1and3} and \eqref{syst_i_for_1and3} have similar forms, the procedure we detailed in the previous section can actually be carried out to deduce all the $h_{a,j}^i$'s. Indeed, with the same steps, we may write the equation satisfied by any $h_{a,j}^i$ as
\begin{equation}
\mathcal{L}[h_{a,j}^i] = \mathcal{S}_{a,j}^i,
\label{eqn_hiaj}
\end{equation}
which is the same equation \eqref{EquDiff_h010} satisfied by $h_{1,0}^0$, but with a non zero right hand side. In the following we may call $\mathcal{S}_{a,j}^i$  `the source term' for $h_{a,j}^i$, because it does not depend on $h_{a,j}^i$. To begin with, let us assume that we know $\mathcal{S}_{a,j}^i$. Then solving \eqref{eqn_hiaj} is straightforward, because the solution of the homogeneous equation $\mathcal{L}[h_{a,j}^i] = 0$ is known (it is a linear combination of $y_P$ and $y_Q$), and we may use the method of variation of parameters to solve it. In fact, apart from $(i,a,j)=(0,1,0)$ and $(i,a,j)=(0,3,1)$, the solution of the homogeneous equation is always equal to zero because the conditions \eqref{H_piSur2} and \eqref{Hprime_piSur2} give $h_{a,j}^i \left(\pi/2\right) = 0 \ \mathrm{and} \ (h_{a,j}^i)'\left(\pi/2\right) = 0$ and therefore the coefficients in the linear combination of $y_P$ and $y_Q$ both vanish. Hence the solution of \eqref{eqn_hiaj} is \citep{BenderOrszag78}
\begin{equation}
\begin{array}{l}
\displaystyle
h_{a,j}^i(\theta) \ = \Gamma(\tfrac{2}{\gamma}) \ y_P(\theta) \int_{\frac{\pi}{2}}^\theta s^{\frac{2}{\gamma}-1} \ y_Q(\theta_1) \ S_{a,j}^i(\theta_1) \ \dd\theta_1 \\
\displaystyle
\hspace{1.2cm} - \Gamma(\tfrac{2}{\gamma}) \ y_Q(\theta) \int_{\frac{\pi}{2}}^\theta s^{\frac{2}{\gamma}-1} \ y_P(\theta_1) \ S_{a,j}^i(\theta_1) \ \dd\theta_1
\end{array}
\label{Solution_hiaj}
\end{equation}
where $\Gamma$ is the Gamma function, $y_P$ and $y_Q$ are defined in \eqref{Def:yP_yQ_InAppendix}, and we used their Wronskian $W[y_P(\theta),y_Q(\theta)]~=~-\sin(\theta)^{1-2/\gamma} / \Gamma(2/\gamma)$.

Let us now illustrate the general procedure to obtain all the source terms $S_{a,j}^i$, by focusing on the special case $\mathcal{S}_{1,0}^1$. We take this as example because it is all that we need to obtain a parabolic expression for $\bar{\omega}_0^2$ (i.e. the expansion up to the second order in $\bar{k}_y$, cf.\ eqn. \ref{omega02_formula_2ndOrder}).
Following the procedure detailed in Appendix \ref{section:SolvingForh01Andh03}, it is straightforward to show that
\begin{equation}
\mathcal{S}_{1,0}^1 = -\frac{2}{\gamma} s^{-\frac{2}{\gamma}} \left[\frac{c}{s} (h_{1,0}^0)'+h_{1,0}^0\right].
\label{S110}
\end{equation}
The point is to notice that this expression (useful for $h_{1,0}^1$) involves a $h_{a,j}^i$ of lower $i$ and $j$ (namely $h_{1,0}^0$), the full expression of which is known (expression \ref{h010}). Therefore with \eqref{h010} and \eqref{S110} we have $\mathcal{S}_{1,0}^1$ explicitly. Hence, with \eqref{Solution_hiaj}, the explicit expression of $h_{1,0}^1$ shown in \eqref{h010_and_h110}.

This feature is general: the explicit expression of any $\mathcal{S}_{a,j}^i$ is always given by some $h_{a,j}^i$ of lower $i$ and $j$, the expression of which is known by iteration, starting from the solution \eqref{h010} of $h_{1,0}^0$. Hence, we may obtain all the $h_{a,j}^i$'s as a sequence of equations that we may solve exactly.

\subsection{Functions $a_0$ and $a_1$}

Now that we have all the $h_{a,j}^i$'s, it is just a matter of bookkeeping to obtain $a_0$ and $a_1$ as expansions in $\bar{k}_y$, i.e. to obtain the coefficients $a_{0,m}$ and $a_{1,m}$ of their respective expansions \eqref{Def:hiaj_a0m_a1m}, by using the definition \eqref{ak} with the $q_a^i$'s directly obtained in terms of the $h_{a,j}^i$'s from \eqref{Def:qa}. The first coefficients are shown in \eqref{a01_a10} and \eqref{a02_a11}.

\bibliographystyle{jfm}
\bibliography{SelfGravFrag}
%
\end{document}